\def\kms{\hbox{km s$^{-1}$}}

\def\hii{H{\sc ii}\,}

\def\msun{M$_{\odot}$\,}

\def\mjyb{mJy beam$^{-1}$}

\def\cm2{cm$^{-2}$}
\def\cm3{cm$^{-3}$}

\def\Ha{H$\alpha$}
\def\radec{\hbox{RA, Dec.(J2000)}}

\def\gra{$^{\circ}$}

\documentclass[printer]{aa}   

\usepackage{natbib}
\usepackage{graphicx}
\usepackage{txfonts}
\begin{document}
\title{  870 $\mu$m continuum  observations of  the bubble-shaped  nebula \hbox{Gum 31} }

\author{N. U. Duronea\inst{1,2}
        \and J. Vasquez\inst{1,2}
        \and L. G\'omez\inst{3,5}
        \and C. E. Cappa\inst{1,2}
        \and V. Firpo\inst{4}
        \and  C. H. L\'opez-Carballo\inst{4}
        \and M. Rubio\inst{5}
        }

\institute{Instituto Argentino de Radioastronom{\'{\i}}a, CONICET, CCT-La Plata, C.C.5., 1894, Villa Elisa, Argentina\ \email{duronea@iar.unlp.edu.ar} \and Facultad de Ciencias Astron\'omicas y Geof{\'{\i}}sicas, Universidad Nacional de La Plata, Paseo del Bosque s/n, 1900 La Plata,  Argentina\and CSIRO Astronomy and Space Science, PO Box 76, NSW 1710 Epping, Australia    \and Departamento de F\'{\i}sica y Astronom\'ia, Universidad de La Serena, Av. Juan Cisternas 1200 Norte, La Serena, Chile \and  Departamento de Astronom{\'{\i}}a, Universidad de Chile, Casilla 36, Santiago de Chile, Chile}

\date{Received: August 2014 / Accepted: June 2015} 
 
 
\abstract
  {}
   { We are presenting here a study of the cold dust  in the close environs of the ring nebula Gum 31. We aim at deriving  the physical properties of the molecular gas and dust  associated with the nebula, and   investigating its correlation with the star formation in the region, that was probably  triggered by the expansion of the ionization front against its environment.    }
   {We make use of 870 $\mu$m emission data obtained with the Large APEX Bolometer Camera (LABOCA) to map the dust emission. The 870 $\mu$m emission provides an excellent probe of mass and density of dense molecular clouds.  The obtained LABOCA image was  compared to   archival infrared, radio continuum, and  optical images.
}
   {The 870 $\mu$m emission follows the 8 $\mu$m {\em (Spitzer)}, 250 $\mu$m, and 500 $\mu$m  ({\em Herschel}) emission distributions showing the classical morphology  of a two dimensional projection of a spherical shell. We use the  870 $\mu$m and 250 $\mu$m images to identify 60 dust clumps in the collected layers of molecular gas using the \texttt{Gaussclumps} algorithm. The clumps have  effective deconvolved radii between 0.16 pc and 1.35 pc, masses between 70 \msun\ and 2800 \msun, and volume densities between  1.1 $\times$ 10$^3$ cm$^{-3}$  and $\sim$ 2.04 $\times$ 10$^5$ cm$^{-3}$. The total mass of the clumps is $\sim$ 37600 \msun. The dust temperature of the clumps is in the range from 21 K to 32 K, while inside the \hii\ region reaches $\sim$ 40 K. The clump mass distribution for the sample is well-fitted by a power law  $d$N/$d$log($M$/M$_{\odot}$) $\propto$ $M^{-\alpha}$, with  $\alpha$ = 0.93 $\pm$ 0.28.  The slope differs from those obtained for the stellar IMF in the solar neighborhood, suggesting that the clumps are not direct progenitors of single stars/protostars. The mass-radius relationship for the 41 clumps  detected in the 870 $\mu$m emission shows that only 37$\%$ of them  lie in or above the high-mass star formation threshold,  most of them having candidate YSOs projected inside their limits.  A comparison of the dynamical age of the \hii\ region with the fragmentation time,   allowed us to conclude that the collect and collapse mechanism may be important for the star formation at the edge of Gum 31, although other  processes may also be acting.  The position of the identified young stellar objects in the region is also a strong indicator that the collect and collapse process is acting. 
}
    {}


\maketitle

\section{Introduction}

There is now consensus that the formation of stars can be triggered by the action of \hii regions over its parental molecular environment.  Processes that sweep up and compress  the gas and dust, like the {\it collect-and-collapse} mechanism (C$\&$C; \citealt{elm77}) and the {\it radiative driven implosion} process (RDI;  \citealt{lela94}),  may  favor the triggering of star formation. Dense molecular condensations (or clumps) lying at the border of  Galactic \hii regions are then among the most likely sites for stellar births, and hence, to look for early stages of star formation (e.g. \citealt{ro09,ca09,va12,deh12,du14}).  In recent years, the triggered star formation process, especially the C$\&$C mechanism, has been studied extensively on the edges of many bubble-shaped  \hii\ regions such as Sh2-104, RCW 79, Sh2-212, RCW 120, Sh2-217, Sh2-90 \citep{deh03,deh08,deh09,zav06,zav10,bra11,sam14}, as well as in several surveys (e.g \citealt{tho12,simp12})

Gum 31 is one of the many southern  Galactic \hii\ regions that appear as {\bf a} ring at infrared (IR) wavelengths (see the upper panel of Fig. \ref{glimpse-870}). The nebula, located at a distance of $\sim$ 2.5 kpc \citep{yon05}, is considered to be part of the giant Carina Nebula Complex \citep{preib12}, and   is ionized by the stars HD 92206A, B, and C, which are members of the young stellar cluster NGC 3324 \citep{jef63,for76,wal82,baum00,maap04}. The  simple morphology of Gum 31, along with the strong evidence of star formation in its environs  (see below),  make this object an excellent laboratory to investigate possible  scenarios of triggered star formation.   Cappa et al. (2008;  hereafter CNAV08) analyzed the  ionized, neutral, and molecular gas in the environs of Gum 31. Adopting a distance of 3 kpc,   CNAV08 found an H{\sc i} shell of $\sim$ 1500 \msun\  around the nebula with a systemic velocity of $-$23 \kms\ expanding at $\sim$ 11 \kms.  The authors also found an associated molecular envelope ($\sim$ 1.1 $\times$ 10$^5$  \msun) at velocities between $-$27.2 \kms\  and $-$14.0 \kms\ that   has been probably accumulated by the expansion of the ionized gas. The distribution of the molecular and  ionized gas, along with the emission of the 8 $\mu$m-MSX band A,  which include significant emission from polyciclic aromatic hydrocarbons (PAHs),   suggests that a photodissociation region (PDR) was created at the interface between the ionization front and the molecular cloud. Using MSX, IRAS, and 2MASS photometric data,  CNAV08  also found a number of infrared young stellar object (YSO) candidates around the nebula which suggests that star formation is active in the high density gas. As part of the  Census of High- and Medium-mass Protostars (CHaMP) program,  \citet{bar11} surveyed the southern sky to detect massive molecular clumps using Mopra (HPBW = 40$''$)   \hbox{HCO$^+$(1-0)} observations. The authors identified 19 molecular clumps in the region of  Gum 31 (see Fig. 34 of that work). The systemic velocity of these clumps is in agreement with the velocity of the molecular gas reported by CNAV08.  Ohlendorf et al. (2013;  hereafter OPG13) used high sensitivity and spatial resolution WISE, {\it Herschel}, and {\it Spitzer} data to characterize the young stellar population surrounding the \hii region. The authors estimated a total population of $\sim$ 5000 YSO candidates in the region, many of them observed in compact clusters and located at the inner edge of the bubble. They concluded that probably the C$\&$C  and  RDI processes are taking place simultaneously at the border of the \hii\ region. The authors based their conclusion  on the size of Gum 31, that agrees with what is expected for a C$\&$C scenario \citep{wi94}, and the presence of YSOs in its rims, which give support to the C$\&$C process, while the existence of YSOs in the head of pillars points to a RDI process. More recently, \citet{vaz14}  analyzed the distribution of the molecular gas and cold dust in the dense clump linked to IRAS 10361-5830, located in the southwestern edge of Gum 31. The authors found a dense molecular shell at $\sim$ $-$25 \kms\ centered at the IRAS position, where star formation is active.  They also found a number of cold dust clumps associated with the molecular shell.

\begin{figure*}
\centering
\includegraphics[width=330pt]{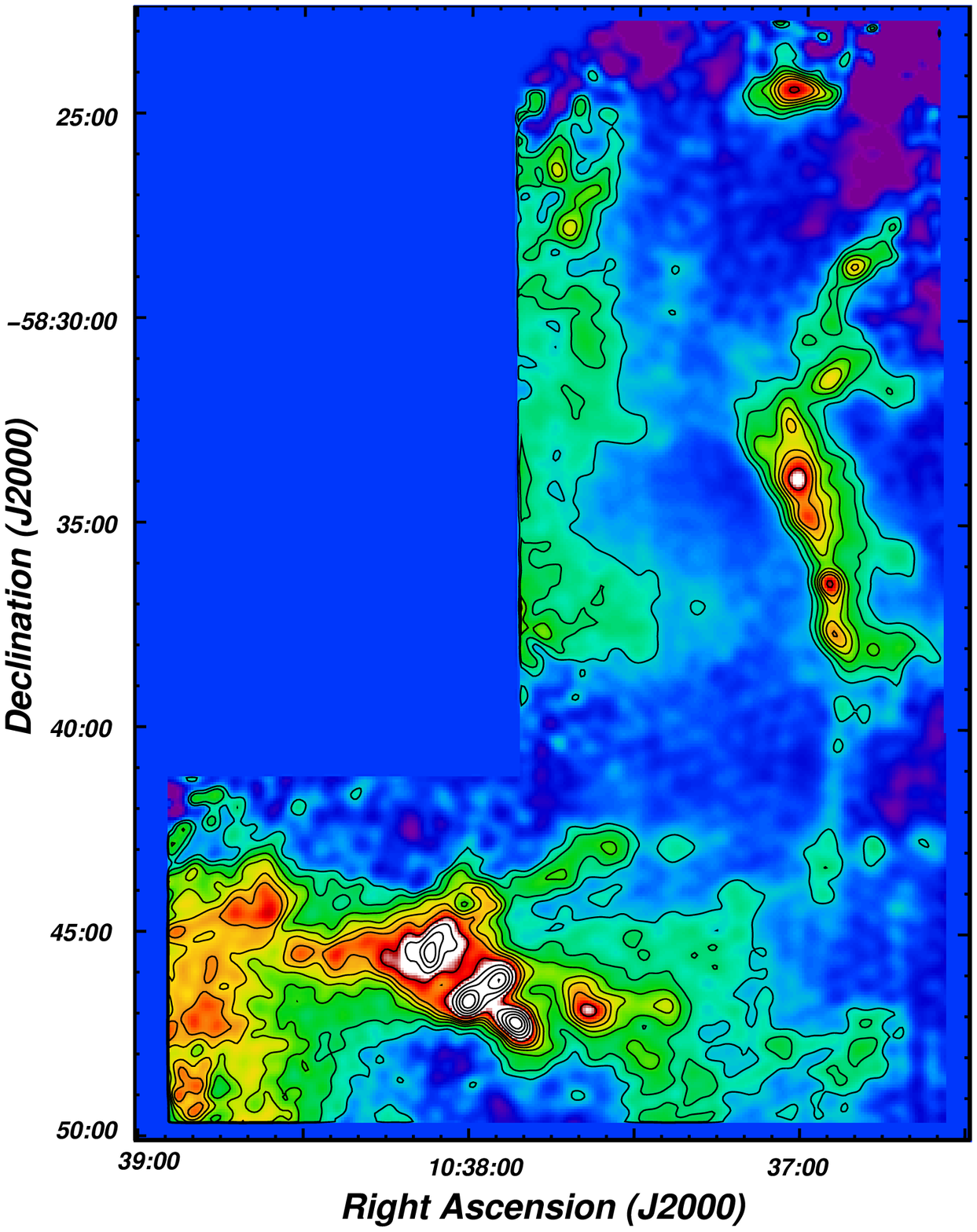}
\caption{ 870 $\mu$m emission of Gum 31. The contours go from  150 \mjyb\ ($\sim$ 3$\sigma$) to 650 \mjyb\ in steps of 100 \mjyb, and from 650 \mjyb\ in steps of 200 \mjyb. The maximum flux density is 1860 \mjyb.  }
\label{870}
\end{figure*}

Although a study of the interstellar medium (ISM) and  the young stellar population around  the whole  Gum\,31 nebula was performed by CNAV08, \citet{bar11}, and OPG13, a high spatial resolution analysis of the densest molecular gas adjacent to the edge of the \hii region still remains to be done.  Low-J transitions of $^{12}$CO and $^{13}$CO are commonly used to trace molecular clouds; however, they are often optically thick, thus probing only the outer layers of the cloud. Morover, also  optically thin transitions from C$^{17}$O and C$^{18}$O  may fail to probe the inner densest molecular gas because they freeze out onto dust grains and deplete at  high densities \citep{her11,gia14}. In this context, optically thin dust continuum emission in the (sub)millimeter range is one of the most reliable tracers of the dense molecular gas from which the stars form.  It provides a powerful tool to probe some basic physical properties of dense clouds (column density of molecular hydrogen, mass of the cloud, etc.) which are needed to unveil the physical conditions in  regions where stars can form. 

 When an \hii\ region expands, molecular gas and dust accumulates behind  the ionization front, forming shells of dense molecular gas  surrounding the ionized gas. With time these shells become massive \citep{hoso06} and could contain cold dust that radiates in the \hbox{(sub-)millimiter} range.  In the present study we analyze the distribution and physical properties of the dense gas and dust adjacent to the ionization front of  Gum 31  using 870 $\mu$m  continuum emission obtained with LABOCA at APEX 12m telescope. The 870 $\mu$m image is compared to IR, radio continuum, and optical archival images  in order to perform a complete multiwavelength study of the dust and  gas associated with the nebula and to  derive their physical properties and conditions.    Optically  thin dust emission at 870 $\mu$m is usually dominated by the thermal emission from cold dust, which is contained in dense material (e.g. dense star-forming cores or filaments).


\section{Observations and complementary data}

The 870 $\mu$m continuum observations were carried out on October 2011 with  the Large Apex BOlometer CAmera (LABOCA)\footnote{APEX is a collaboration between the Max-Planck-Institut f\"ur Radioastronomie, the European Southern Observatory, and the Onsala Space Observatory}. LABOCA is a 295-pixel bolometer array developed by the Max-Planck-Institut f\"ur Radioastronomie \citep{sir07}. The central frequency of the instrument is 345 GHz and  the bandwidth is 60 GHz. The beam size  (HPBW) at 345 GHz is 19\farcs 2 (or 0.23 pc at 2.5 kpc). Observations were made using the on-the-fly (OTF) mode to map  an inverted ``L''-shaped region   of the nebula where  the 8 $\mu$m emission is more intense. During the observations the amount of precipitable water vapour (PWV) was between 0.16 mm and 0.24 mm.    Absolute flux calibrations were achieved through observations of the planet Mars as primary calibrator for LABOCA, and the stars N207\,11R and  VY\,CMa as secondary calibrators. The uncertainty due to flux calibration was estimated to be $\sim$ 20\%. The telescope focus and pointing were checked using the star $\eta$ Carinae.   Observations were smoothed down to a beam of 20\farcs0 to obtain a final average rms noise of 50 mJy beam$^{-1}$. The data were reduced  using the {\it Comprehensive Reduction Utility for SHARC-2 software package} (CRUSH-2)\footnote{http://www.submm.caltech.edu/$\sim$sharc/crush/index.html} \citep{kov08} following the standard procedure.  The 870 $\mu$m emission image of the  Gum 31 is shown in Fig \ref{870}.

The 870 $\mu$m data were complemented with several archival data sets:
\begin{itemize}

\item  Infrared  images from the  the {\em Herschel}\footnote{{\em Herschel} is an ESA space observatory with science instruments  provided by European-led Principal Investigator consortia and with important participation from NASA (http://www.cosmos.esa.int/web/herschel/science-archive)} Infrared GALactic (Hi-GAL) plane survey key program \citep{mol10}. We used the 70\,$\mu$m and 160\,$\mu$m maps produced in the High-Level 2.5, and 250\,$\mu$m and 500\,$\mu$m maps produced in the High-Level 3 (combined maps from Parallel Mode observations done in both nominal and orthogonal scan directions) available in the Herschel Legacy as a standalone product. The Level 2.5 maps were produced with correctedmadmap in the MADMap software application \citep{cant10} with point source artifacts corrected. The pixel sizes of the two PACS maps are 3\farcs 2 (for the 70\,$\mu$m map) and 4\farcs 5 (for the 160 \,$\mu$m map), as suggested in \citet{traf11}. The angular resolution for 70\,$\mu$m and 160\,$\mu$m bands are 10\arcsec and 13\farcs 5, respectively. The Level 3 were produced from the Planck zero-point calibrated maps and the angular resolution for 250\,$\mu$m and 500\,$\mu$m maps are 18\arcsec and 36\arcsec. The {\em Herschel} Interactive Processing Environment (HIPE v12\footnote{HIPE is a joint development by the Herschel Science Ground Segment Consortium, consisting of ESA, the NASA Herschel Science Center, and the HIFI, PACS and SPIRE consortia  members, see http://herschel.esac.esa.int/HerschelPeople.shtml}; \citealt{ott10}) was used to handle the maps.\\

\item  Infrared Spitzer images at 8.0 $\mu$m with a spatial resolution of $\sim$ 2\arcsec, retrieved from the Galactic Legacy Infrared Mid-Plane Survey Extraordinaire (GLIMPSE)\footnote{ http://sha.ipac.caltech.edu/applications/Spitzer/SHA} \citep{b03}.\\

\item Radiocontinuum data from the survey of the Sydney University Molonglo Sky Survey (SUMSS)\footnote{http://www.astrop.physics.usyd.edu.au/cgi-bin/postage.pl} \citep{b99} at 843 MHz.\\

\item Optical data from the 2nd Digitized Sky Survey  (red plate)\footnote{http://skyview.gsfc.nasa.gov/cgi-bin/query.pl}  \citep{mcl00}.

\end{itemize}

\section{Observational results and discussion}

\subsection{870 $\mu$m emission and comparison with other wavelengths}

 In Fig. \ref{870} we show the 870 $\mu$m emission of the Gum 31 nebula.  Despite the difference in spatial resolution, the 870 $\mu$m emission coarsely follows the CO(1-0) emission distribution depicted in CNAV08 (see their  Fig. 5).  The emission at this wavelength  shows the typical IR morphology of Galactic ring nebulae \citep{chu06,chu07,deh10} with a very intense emission towards the border of the nebula and a very faint emission towards its center.  From the figure, we note that the dust emission in the border is not uniformly distributed, and several condensations can be distinguished  in the whole structure. These condensations will be identified and analyzed in the next sections.    

 The ring morphology in the IR  emission of many  \hii\ regions  is a common feature in the Galaxy (e.g. \citealt{wat08,deh09,deh10}) and has originated some debate on whether these structures are flat (few parsecs) two-dimensional ring-like objects \citep{bea10} or three-dimensional bubbles \citep{deh10,and12,and15}.  Although it is difficult to  distinguish between the two different scenarios, the observational properties suggest that the ring morphology of Gum 31 in the IR  may be due to  a projection effect, namely: 1)  A  weak  structure is observed in the 870 $\mu$m emission  towards the center of the nebula along RA $\approx$  10$^h$ 37$^m$ 40$^s$  (see Fig. \ref{870}).  This structure probably corresponds to IR emission of cold dust  in  front of the nebula, since its outer border shows an excellent spatial correlation with  a region of optical absorption and 70 $\mu$m emission (see upper panel of Fig. \ref{glimpse-870}). The  optically-absorbed region  lies at  RA $>$  10$^h$ 37$^m$ 45$^s$, which lead us to conclude that probably there is more  870 $\mu$m emission, arising from the front face of the bubble in the not-covered area with LABOCA. 2)   The velocity interval  of the high-density molecular gas adjacent to the nebula ($\sim$ $-$16 \kms; \citealt{bar11}), which indicates the presence of  expanding motions that are not suitable to flat face-on structures.   3) The optical and radio continuum emission distribution inside the nebula (see below), which resembles a filled sphere of  ionized gas inside the IR emission.  
\begin{figure*}
\centering
\includegraphics[width=430pt]{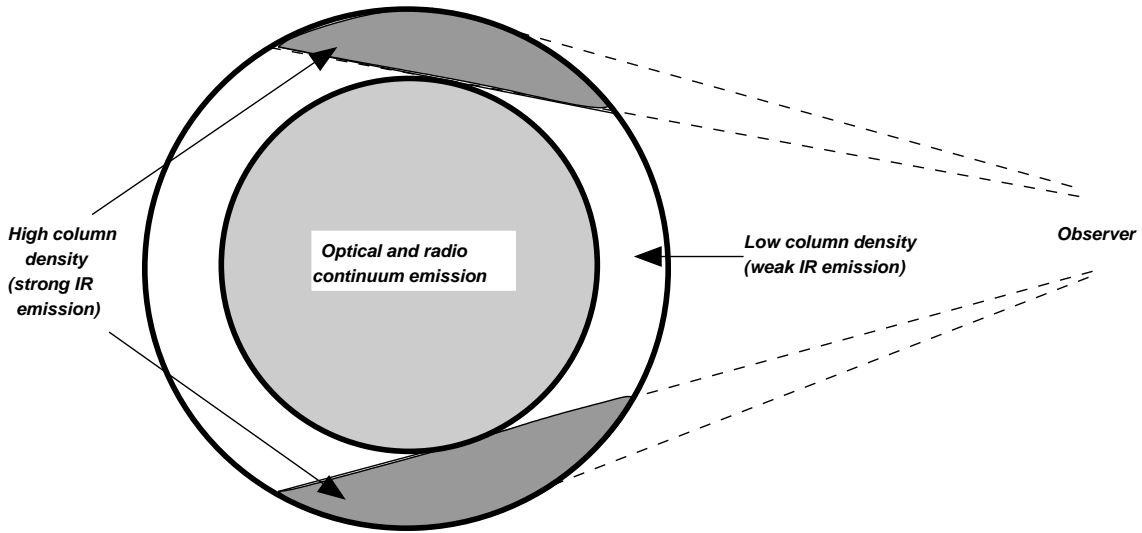}
\caption{ Sketch of Gum 31, which could explain  in a qualitative way the IR, optical and radio continuum emission distribution.   }
\label{esquema}
\end{figure*}

 In a qualitative manner, the morphology of Gum 31 (strong IR emission towards the border of the  nebula  and weak emission at the center)  depicts the typical emission distribution of optically thin emission of a bubble, scaling  with the line-of-sight path length towards the edge of the shell. This effect is known as ``limb brightening''  (see Fig.~\ref{esquema}).  The weak 870 $\mu$m emission at the center may be due to gas/dust from the front face of the bubble  with too low column density to be firmly detected.  The limb brightening effect has  been used to explain the shape of many IR ring bubbles (e.g. \citealt{wat08,deh10}), and could be also used to explain the spatial distribution of candidate YSOs observed in the nebula, which seems to be more crowded at the edge of the bubble  (see Fig. 8 of OPG13). High resolution molecular observations  of the whole nebula would be useful to confirm the bubble morphology proposed above for Gum 31.

\begin{figure*}
\centering
\includegraphics[width=330pt]{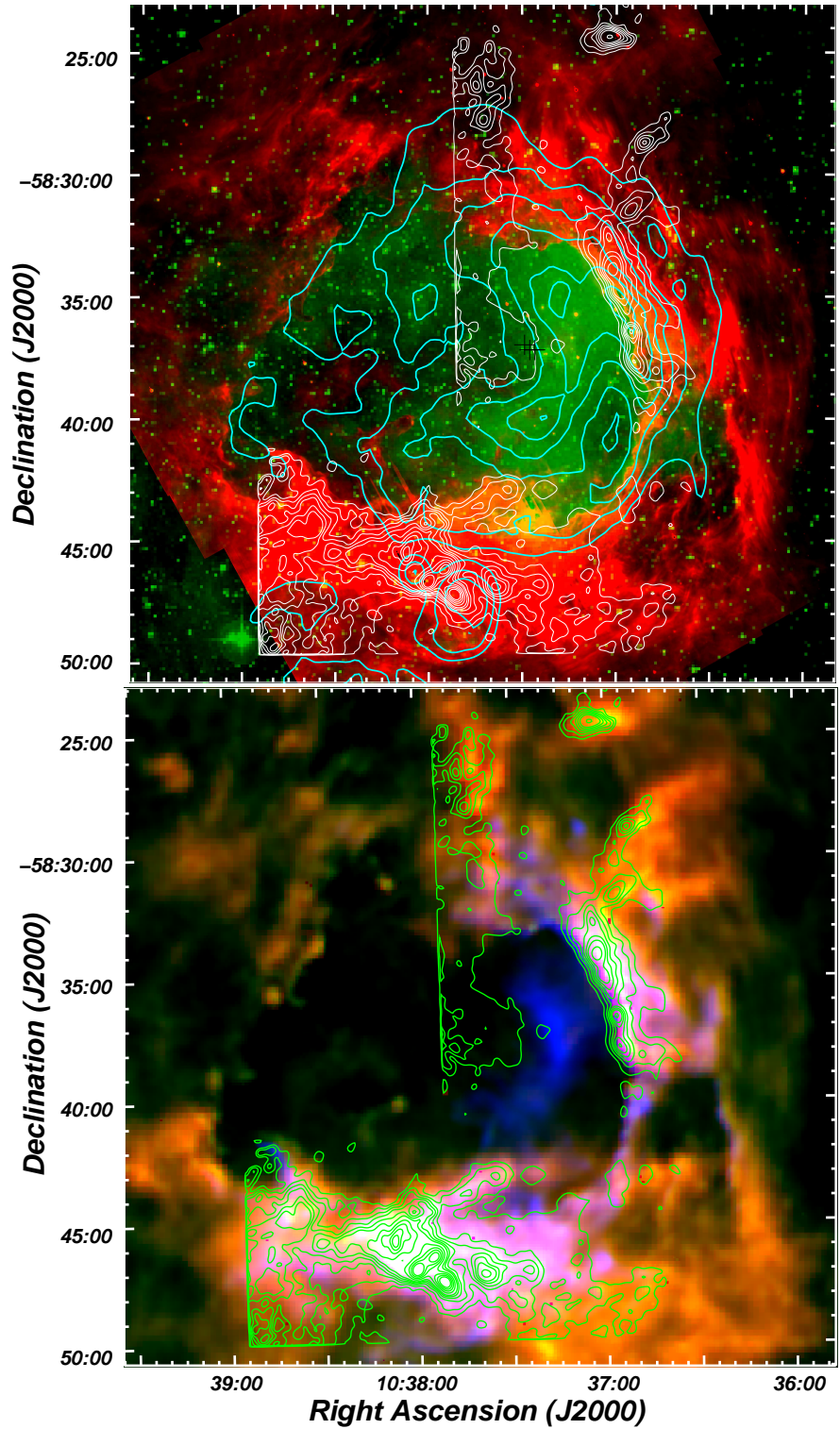}
\caption{ {\it Upper panel:} Composite image of Gum 31 and its environs. The IRAC-GLIMPSE emission at 8.0 $\mu$m is shown in red, while green  indicates the  DSSR2 optical emission (red plate). White contours show the 870 $\mu$m emission while cyan contours shows the radio continuum emission at 843 MHz. Black crosses at the center  show the position of stars HD 92206A, B, and C. {\it Lower panel:} Composite image of the 870 $\mu$m dust continuum emission (green contours) overlaid onto the  {\it Herschel} 70  $\mu$m (blue), 250 $\mu$m (green),  and 500 $\mu$m (red) emissions. The contours corresponding to the 870 $\mu$m emission in both panels are the same as in Fig.\ref{870}.  }
\label{glimpse-870}
\end{figure*}

 In the upper  panel  of Fig. \ref{glimpse-870} we show  the 870 $\mu$m emission (white contours)  and the SUMSS 843 MHz radio continuum emission  (cyan contours) overlaid onto the optical (DSSR2)  image (in green)  and infrared (IRAC-GLIMPSE)  8 $\mu$m emission (in red). At 8 $\mu$m, most of the emission originates in strong features of PAH molecules,  which are considered to be good tracers of warm UV-irradiated photodissociation regions (PDR; \citealt{ht97}).  Since these complex molecules are destroyed inside the ionized gas of an \hii\ region \citep{ces96,pov07,leb07}, they indicate the limits of the ionization front and delineate the boundaries of the bubble nebula. The 870 $\mu$m emission follows the 8 $\mu$m emission distribution, which  is bright in the western and southern regions of Gum 31. The 843 MHz radio continuum and optical emissions appear mostly confined {\bf within} the IR bubble. Three emission peaks  can be noticed in the radio continuum emission  at \radec\ =\ 10$^h$37$^m$24$^s$, $-$58\gra39\arcmin57\arcsec, \radec\ =\ 10$^h$37$^m$18$^s$, $-$58\gra34\arcmin50\arcsec, and \radec\ $\approx$ 10$^h$36$^m$59$^s$, $-$58\gra35\arcmin20\arcsec. The last one is elongated and coincident with strong  870 $\mu$m emission, which  strongly suggests that dense cold dust and molecular gas have  been compressed and collected behind the ionization front due to the expansion of the \hii region. The other two radio continuum maxima are coincident with strong optical emission.

 The presence of intense emission   in optical, 8 $\mu$m, radio continuum,  and 870 $\mu$m emission,  to the west of the powering stars speaks in favor of  the existence of a PDR  at the interface between the ionized and molecular gas, as suggested by CNAV08.  On the other hand,  the strong  radio continuum, optical,  and IR emission, observed in the western region of the bubble in comparison with the emission at the eastern and northeastern edges  suggests that the  \hii\ region is expanding  anisotropically, probably due to the lack of dense molecular gas at the eastern and northeastern edges of the bubble.

 As can be seen in Fig. \ref{glimpse-870}, a  considerable deficit in the 8 $\mu$m and 870 $\mu$m emission is  visible around   \radec\ $\approx$ 10$^h$36$^m$50$^s$, $-$58\gra43\arcmin20\arcsec, coincident with a lack of CO emission (CNAV08) and  with the presence of  some weak \Ha\ and radio continuum emission located outside the border of the PDR. In Sect. 3.3 (Fig. \ref{fig:tdust})   we will show that this region is  correlated with a significant dust temperature enhancement. This could indicate that UV photons from the nebula could be leaking into the ISM through a discontinuity in the PDR.  Two other interesting features can be noticed when comparing the  8 $\mu$m and 870 $\mu$m emissions at \radec\ $\approx$ 10$^h$37$^m$03$^s$, $-$58\gra24\arcmin23\arcsec  and  \radec\ $\approx$ 10$^h$37$^m$42$^s$, $-$58\gra27\arcmin00\arcsec. The 870 $\mu$m structures at these positions (which will be identified in Sect. 3.2 as clumps 8, 18, 19, and 35) appear projected onto deep spots of absorption at  8 $\mu$m emission, a typical feature of the so-called infrared dark clouds (IRDCs; \citealt{egan98,rath06}). 

 In the lower panel of Fig. \ref{glimpse-870} we show  the 870  $\mu$m emission (green contours)  overlaid on the 3 color image of the {\it Herschel} 70 $\mu$m (in blue), 250 $\mu$m (in green), and 500 $\mu$m (in red)    emissions.  The  250 $\mu$m and 500 $\mu$m emission follow the 870 $\mu$m emission, although they seem to be a bit more extended. This means that the {\it Herschel} emission  at high wavelengths is still good tracer of cold dust in Gum 31.  A comparison  between the 870 $\mu$m emission and the 250 $\mu$m and 500 $\mu$m emissions in the southern part of the nebula (Dec $\sim$  $-$58\gra 46\arcmin 00\arcsec)   shows that the area mapped with LABOCA almost covers the whole cold dust in the region.    Since the emission at    250 $\mu$m and 500 $\mu$m  is extremely faint at the eastern and north-eastern  regions of Gum 31, where no CO(1-0) emission  was detected either  (\citealt{yon05}, CNAV08), we infer that there is  little dense molecular gas/dust  in  the not-mapped area of the \hii\ region (see  Sect. 3.3.4). This may indicate that Gum 31 has evolved in an  inhomogeneous medium, with  more material and higher densities towards its  western edge.  The 70 $\mu$m emission, which traces warm dust, is very intense along a very thin lane at the inner border of the PDR. This is a common feature observed in many Galactic IR bubbles (e.g   \citealt{and12}). The 70 $\mu$m emission is also  coincident with strong 870 $\mu$m emission in the southern region of the bubble,  where the emission at  8 $\mu$m, 250 $\mu$m, and 500 $\mu$m is also intense; this gives an idea about the complex nebula morphology.

\subsection{Identification of dust clumps at 870 $\mu$m}

   As mentioned in Sect. 3.1, Figure \ref{870} shows  that the dust emission is not uniformly distributed, and several condensations can be distinguished  in the whole structure.   These condensations must be identified and their physical properties must be estimated if a study of the physical properties of the cold dust and molecular gas in the whole  bubble nebula  is to be carried out. Different terminologies have been used in the last years to describe the substructure of a  molecular cloud. A hierarchical categorization  was proposed by  \citet{blitz93}, for  structures of different scales in the interstellar medium, as: {\it clouds}, {\it clumps}, and {\it cores}. Clouds have masses $>$ 10$^4$ \msun, and cores (regions where single stars or binary systems might be formed) have masses $<$ 10  \msun and sizes $<$ 0.1 pc. Clumps are then structures with intermediate masses and sizes from which stellar clusters can be formed.  A similar classification was also proposed by \citet{ber07} and  \citet{will2000} designating  individual overdensities inside a molecular cloud, which can be identified by different automated algorithms, as clumps.  Given the characteristics mentioned above and the physical properties derived for the condensations distinguished in the 870 $\mu$m emission, we will  hereafter refer  to these structures as ``clumps''. 

 We have identified the dust clumps  in Gum 31 in a systematic way with the commonly used \texttt{Gaussclumps} algorithm \citep{stu90,kra98}, which is      a task included in the GILDAS\footnote{http://www.iram.fr/IRAMFR/GILDAS}   package. Although  \texttt{Gaussclumps}  was originally written to work on  3-dimensional data cube, it can be modified to be applied to continuum images without modification of the code by adding two 2-dimensional empty planes to the original continuum image mimicking a 3-dimensional datacube (see \citealt{mote03}). \texttt{Gaussclumps}  uses a least-squares fitting procedure to decompose the image/cube into a series of Gaussian-shaped clumps by iteratively substracting fitted clumps.   The algorithm   identifies  the position, peak flux density, full width at half  maximum (FWHM), and position angle of the fitted sources. As suggested by \citet{kra98}, the ``stiffness'' parameters that control the fitting were all set to 1. The peak flux density threshold was set to 4$\sigma$.
\begin{figure}
\centering
\includegraphics[width=260pt]{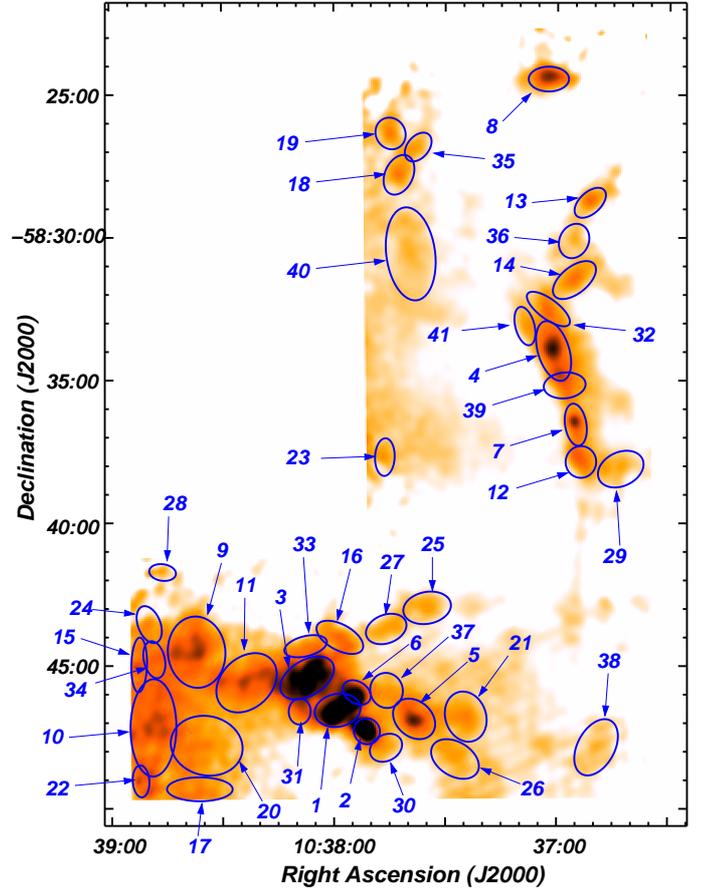}
\caption{ Clumps identified with  \texttt{Gaussclumps}  in the  870 $\mu$m emission image.    }
\label{gaussclumps}
\end{figure}
\begin{figure}
\centering
\includegraphics[width=240pt]{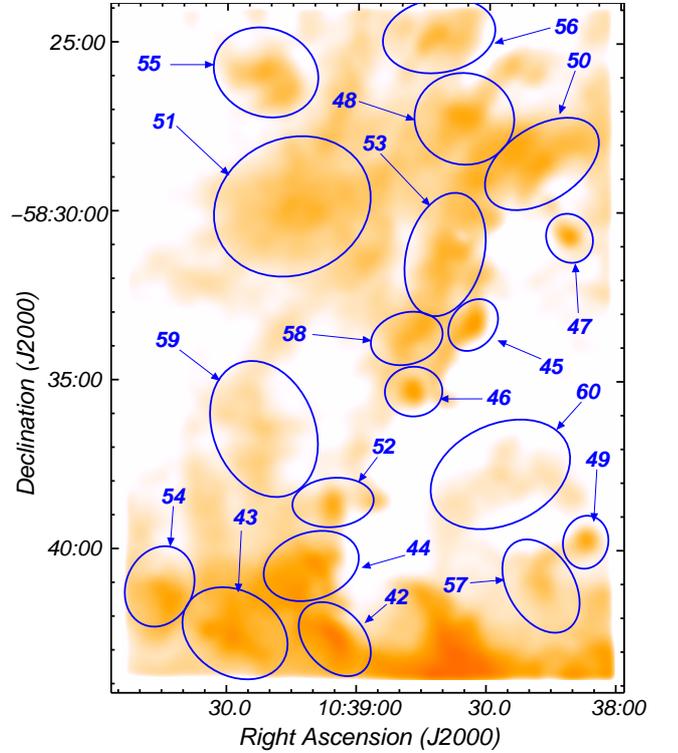}
\caption{ Clumps identified with \texttt{Gaussclumps} algorithm  in the  250 $\mu$m emission image,  over the not-covered region in 870 $\mu$m emission.    }
\label{gaussclumps-250}
\end{figure}
 To avoid false detections, we have trimmed the noisy borders of the image  and   added a wedge of null values to force  \texttt{Gaussclumps} to identify clumps inside the image.  To identify  genuine clumps  we have imposed that the deconvolved size  of the clumps should be larger than 50 $\%$ of the HPBW size (i.e. $R_{\rm D}$ $>$ HPBW/2, or 0.12 pc at a distance of 2.5 kpc).  As a result, 41 clumps were identified in the 870 $\mu$m emission image.

  In Table \ref{tabla-prop} we present all the clumps identified with \texttt{Gaussclumps}. The identification number of the clumps  and their coordinates are indicated in Cols. 1 and 2, respectively.  The spatial location, numerical identification, size, and orientation derived for each clump are depicted  in Fig. \ref{gaussclumps}. As can be seen from this figure, most of clumps satisfy a convincing visual identification. In Col. 11 of Table  \ref{tabla-prop} we indicate the HCO$^{+}$ clumps reported by \citet{bar11} that are projected over the corresponding 870 $\mu$m clump.  Clumps  1, 6, and 3 were reported  by \citet{vaz14}, labeled in that work as D1, D2, and D3, respectively (see Fig. 7 of that work).  

  In Sect. 3.1, we pointed out that  clumps 8, 18, 19, and 35 appear projected onto deep spots of absorption at  8 $\mu$m emission, which suggests that these clumps could be foreground dense objects. These clumps, very bright at 250 $\mu$m and 500 $\mu$m but no emitters at 70 $\mu$m, appear quite isolated and projected slightly farther form the nebula, casting some doubts on their association with the rest of the clumps and their interaction with the ionization front. Since their velocities are in agreement with the rest of the clumps \citep{bar11} we infer  that these clumps could be a remnant of the parental cloud that remains undisturbed by the expansion of the nebula.

 With the aim of studying the north-eastern region of the Gum 31 nebula, that was not observed with LABOCA (see Sect. 3.1), we make use of {\em Herschel} images at 250  $\mu$m since the emission at this wavelength is still a good tracer of cold dust. The identified clumps are presented in the lower part of Table \ref{tabla-prop}. The number identification of the clumps are continued from those identified at 870 $\mu$m.  The spatial location, numerical identification, size, and orientation derived for each clump are depicted  in Fig. \ref{gaussclumps-250}.  We have rejected from the sample two clumps around the position  \radec\ $\approx$ 10$^h$38$^m$35$^s$, $-$58\gra 42\arcmin 30\arcsec, since they likely are the 250 $\mu$m counterparts of clumps 9, 24, and 28.  We also note that clumps 49, 57, and  60  are projected over the center of the nebula and   show a good spatial correlation with  regions of optical absorption (see Fig. \ref{glimpse-870}) which could indicate that these clumps are placed in front of the nebula. This clumps have low column and volume densities and could be related with the faint  IR emision at the center of the nebula, giving more support to the  bubble  morphology proposed earlier for Gum 31.

 We are  cautious  about the  clump identification described above, as well as the physical properties (see next section), since  the IR nebula  was proposed to be the result of a  projection of a three dimensional bubble onto a two dimensional plane. This problem is, however, not only concerning to the case of Gum 31, but  should be also addressed in  many other molecular  studies of Galactic  bubble nebulae, where different molecular  clumps (or condensations) were studied.

\begin{table*}
\centering
\caption[]{Properties of identified clumps  around the bubble \hii\ region Gum 31. Clumps 1 to 41 were identified at 870 $\mu$m while clumps 42 to 60 at 250 $\mu$m }
\label{tabla-prop}
\begin{tabular}{cccccccccccc}
\hline
 clump  &  RA, \  Dec$_{\rm J2000}$ & $I_{\rm peak}$   &  $S_{\nu}$ &  $\theta_{\rm maj}\times \theta_{\rm min}$ & $R_{\rm D}$  &  $M_{\rm tot}$ &  $n$   & $N_{\rm H_2}$   &  $T_{\rm dust}$    & HCO$^{+}$   &        \\
number   & ( $^h$\  $^m$\  $^s$, \gra\  \arcmin\  \arcsec ) &  ($\frac{\rm Jy}{\rm beam}$) &   (Jy) &  ( \ \arcsec\ \ $\times$\ \ \arcsec \ )  &  (pc)         & (M$_{\odot}$)  &   (cm$^{-3}$)      & (10$^{22}$cm$^{-2}$) &  (K) & (BYF$^{(\dag)}$   ) &          \\
\hline
\hline
 & & & & & & & & & &  &       \\
           1 &  10 37 59,       -58 46 34  &    1.76   &    23.41  &     86.8   $\times$  61.1  &    0.42   &   2050   &  93500   & 10.4  &     25  &  77b  & \\
           2 &  10 37 51, 	-58 47 18  &    1.71   &     8.21  &     45.7   $\times$  41.8  &    0.23   &   770    &  203800  & 10.9  &   24   &   77a & \\
           3 &  10 38 07, 	-58 45 26  &    1.59   &    30.26  &     113.9  $\times$  66.5  &    0.51   &   2800   &  72300   & 9.9   &  24    &    77c & \\
           4 &  10 37 01, 	-58 33 59  &    0.96   &    16.62  &     120.3  $\times$  57.5  &    0.48   &  1450    &  43800   & 5.6  &   25   &  70a &  \\
           5 &  10 37 38, 	-58 46 52  &    0.92   &    13.51  &     87.5   $\times$  66.7  &    0.44   &  1130    &  43900   & 5.2   &  26   & 77d &  \\
           6 &  10 37 54, 	-58 45 57  &    0.93   &    4.65   &      53.5  $\times$  37.2  &    0.23   &  410     &  103400  & 5.5   &  25   & 77b &  \\
           7 &  10 36 55, 	-58 36 34  &    0.83   &    6.11   &     79.4   $\times$  36.8  &    0.29   &  530     &  70700   &  4.9  &   25   &   70b &  \\
           8 &  10 37 02, 	-58 24 27  &    0.87   &    7.21   &      74.6  $\times$  44.1  &    0.32   &  820     &  84400   &  6.7  &  21   & 67  & \\
           9 &  10 38 37, 	-58 44 32  &    0.79   &    28.03  &     130.1  $\times$  108.7 &    0.71   &  2430    &  23300   &  4.6  &  25   &&  \\
          10 &  10 38 49, 	-58 47 10  &    0.76   &    32.98  &     194.8  $\times$  88.3  &    0.78   &  2640    &  18900   &  4.1  &  26   &&  \\
          11 &  10 38 24, 	-58 45 36  &    0.73   &    18.59  &     132.8  $\times$  76.1  &    0.59   &  1640    &  26900   &  4.4  &  25  &&  \\
          12 &  10 36 54, 	-58 37 52  &    0.56   &    4.52   &     56.7   $\times$  56.1  &    0.31   &  390     &  40600   &  3.2  &  25  & 70b &   \\
          13 &  10 36 51, 	-58 28 47  &    0.56   &    3.65   &     69.3   $\times$  37.1  &    0.27   &  370     &  60800   &  3.9  &  22   &  66 &  \\
          14 &  10 36 56, 	-58 31 30  &    0.52   &    6.10   &     96.8   $\times$  48.2  &    0.39   &  580     &  33600    &  3.3  &  23   &    69  &  \\
          15 &  10 38 53, 	-58 44 58  &    0.54   &    3.56   &     105.0  $\times$  24.8  &    0.23   &  290     &  76200   &  3.0  &  26   &&  \\
          16 &  10 37 58, 	-58 44 01  &    0.51   &    5.84   &     97.4   $\times$  46.2  &    0.38   &  480     &  29600   &  2.9  &  26   &&  \\
          17 &  10 38 36, 	-58 49 20  &    0.51   &    6.95   &     127.8  $\times$  42.3  &    0.41   &  540     &  26100   &  2.7  &  27 &&  \\ 
          18 &  10 37 43, 	-58 27 49  &    0.48   &    4.93   &     77.8   $\times$  52.7  &    0.36   &  470     &  32500   &  3.1  &  23   & 72 &  \\
          19 &  10 37 45, 	-58 26 22  &    0.48   &    3.75   &     58.2   $\times$  53.6  &    0.31   &  370     &  40600   &  3.2  &  23   &72 &  \\
          20 &  10 38 34, 	-58 47 48  &    0.48   &    16.91  &     130.1  $\times$  106.5 &    0.70   &  1290    &  12800   &  2.5  &  28  &&   \\
          21 &  10 37 25, 	-58 46 48  &    0.46   &    8.71   &     97.5   $\times$  77.4  &    0.51   &  700     &  17900   &  2.5  &  27  &&   \\
          22 &  10 38 52, 	-58 49 03  &    0.50   &    1.94   &     57.2   $\times$  26.8  &    0.18   &  150     &  77100   &  2.6  &  27  &&   \\
          23 &  10 37 46, 	-58 37 42  &    0.40   &    2.30   &     69.2   $\times$  33.2  &    0.25   &  140     &  30100   &  1.7  &  32 &&   \\
          24 &  10 38 50, 	-58 43 33  &    0.38   &    2.68   &     71.5   $\times$  39.1  &    0.29   &  200     &  27900   &  1.9  &   28 &&   \\ 
          25 &  10 37 35, 	-58 42 59  &    0.36   &    4.86   &     91.1   $\times$  58.9  &    0.42   &  390     &  17400   &  1.9  &  27  &&  \\ 
          26 &  10 37 27, 	-58 48 17  &    0.34   &    5.48   &     99.0   $\times$  63.7  &    0.46   &  470     &  16300   &  2.0  &   25 &&  \\  
          27 &  10 37 46, 	-58 43 43  &    0.34   &    3.44   &     78.1   $\times$  51.8  &    0.36   &  270     &  19100   &  1.8  &  27  &&  \\ 
          28 &  10 38 46, 	-58 41 44  &    0.33   &    1.05   &     46.5   $\times$  26.8  &    0.16   &  70      &  52700   &  1.5  &  28 &&  \\ 
          29 &  10 36 43, 	-58 38 07  &    0.33   &    4.79   &      90.5  $\times$  63.0  &    0.44   &  390     &  15600   &  1.8  &  26  && \\ 
          30 &  10 37 46, 	-58 47 53  &    0.34   &    2.37   &     61.3   $\times$  44.5  &    0.29   &  170     &  23500   &  1.6  &   29  && \\  
          31 &  10 38 09, 	-58 46 37  &    0.34   &    1.47   &     44.9   $\times$  38.0  &    0.21   &  120     &  38600   &  1.8  &  27  &&  \\ 
          32 &  10 37 02, 	-58 32 31  &    0.32   &    2.68   &     98.3   $\times$  33.7  &    0.30   &  240     &  27900   &  1.9  &  24  &&  \\ 
          33 &  10 38 08, 	-58 44 20  &    0.35   &    2.35   &     83.4   $\times$  32.2  &    0.27   &  180     &  30900   &  1.8  &  27   &&  \\
          34 &  10 38 49, 	-58 44 47  &    0.33   &    2.09   &     69.1   $\times$  36.1  &    0.27   &  170     &  30200   &  1.9  &   26  &&  \\
          35 &  10 37 37, 	-58 26 50  &    0.31   &    1.74   &     59.0   $\times$  37.4  &    0.25   &  160     &  34700   &  2.0  &   24  & 72 & \\
          36 &  10 36 56, 	-58 30 07  &    0.31   &    2.69   &     64.8   $\times$  53.1  &    0.33   &  240     &  22100   &  1.9  &   25  &&  \\
          37 &  10 37 46, 	-58 45 53  &    0.33   &    3.22   &     64.1   $\times$  60.4  &    0.35   &  240     &  17700   &  1.6  &  28  && \\
          38 &  10 36 49, 	-58 47 51  &    0.29   &    6.04   &     117.7  $\times$  70.0  &    0.53   &  520     &  11700   &  1.7  &  25  && \\
          39 &  10 36 58, 	-58 35 11  &    0.30   &    2.75   &      78.0  $\times$  46.2  &    0.34   &  220     &  19200   &  1.6  &  28  && \\
          40 &  10 37 39, 	-58 30 35  &    0.28   &    12.56  &     187.0  $\times$  94.5  &    0.79   &  990     &   6800   &  1.5  &  27  && \\
          41 &  10 37 09, 	-58 33 07  &    0.29   &    1.73   &     72.8   $\times$  32.0  &    0.25   &  130     &  27600   &  1.5  &  28   && \\
& & & & & & & & & & &        \\
\hline
& & & & & & & & & & &        \\ 
42 & 10 39 05, -58 42 40  & 8.94 &   329.38  &  147.3 $\times$  80.9   &    0.65   &   550   &     6800   &    1.2 &  25    && \\ 
43 & 10 39 28, -58 42 36  & 7.31 &   464.92  &  162.3 $\times$ 126.9   &    0.86   &   810   &     4400   &    1.1 &  24    && \\ 
44 & 10 39 10, -58 40 25  & 6.91 &   270.21  &  127.7 $\times$  99.2   &    0.67   &   470   &     5300   &    1.0 &  24    && \\
45 & 10 38 34, -58 33 21  & 6.21 &   153.44  &  101.8 $\times$  78.5   &    0.53   &   240   &     5600   &    0.8 &  25    && \\
46 & 10 38 47, -58 35 20  & 6.04 &   170.64  &  103.5 $\times$  88.4   &    0.56   &   270   &     5100   &    0.8 &  25     && \\
47 & 10 38 11, -58 30 46  & 5.69 &   128.66  &   90.4 $\times$  81.0   &    0.50   &   200   &     5300   &    0.7 &  25     && \\
48 & 10 38 36, -58 27 15  & 5.94 &   394.71  &  156.4 $\times$ 137.6   &    0.88   &   680   &     3400   &    0.8 &  24     && \\
49 & 10 38 07, -58 39 45  & 5.45 &   129.48  &   94.8 $\times$  81.1   &    0.52   &   180   &     4400   &    0.6 &  27     && \\
50 & 10 38 18, -58 28 34  & 5.59 &   397.16  &  208.4 $\times$ 110.5   &    0.91   &   520   &     2400   &    0.6 &  27    && \\
51 & 10 39 15, -58 29 52  & 5.44 &   846.49  &  242.6 $\times$ 207.4   &    1.35   &  1390   &     1900   &    0.7  & 25    && \\
52 & 10 39 06, -58 38 37  & 5.28 &   149.84  &  120.4 $\times$  76.3   &    0.56   &   230   &     4300   &    0.7 &  26    && \\
53 & 10 38 40, -58 31 16  & 5.29 &   368.83  &  178.7 $\times$ 126.2   &    0.90   &   510   &     2400   &    0.6 &  27    && \\
54 & 10 39 45, -58 41 16  & 5.33 &   292.23  &  147.5 $\times$ 120.4   &    0.79   &   520   &     3450   &    0.8 &  24     && \\
55 & 10 39 21, -58 25 54  & 5.20 &   312.22  &  160.3 $\times$ 121.3   &    0.83   &   600   &     3500   &    0.8 &  23     && \\
56 & 10 38 42, -58 24 47  & 5.09 &   285.67  &  184.5 $\times$ 184.5   &    0.80   &   610   &     4000   &    0.9 &  23     && \\
57 & 10 38 17, -58 41 04  & 4.79 &   290.35  &  176.9 $\times$ 111.0   &    0.84   &   360   &     2100   &    0.5 &  28     && \\
58 & 10 38 49, -58 33 44  & 4.69 &   175.54  &  131.5 $\times$ 92.0    &    0.65   &   270   &     3300   &    0.6 &  25     && \\
59 & 10 39 22, -58 36 27  & 4.23 &   597.59  &  251.0 $\times$ 182.3   &    1.29   &   890   &     1400   &    0.5 &  26     && \\
60 & 10 38 27, -58 37 46  & 3.75 &   542.53  &  263.4 $\times$ 177.8   &    1.30   &   670   &     1100   &    0.4 &  28     && \\

& & & & & & & & & & &        \\
\hline
\end{tabular}
\tablefoot{
\tablefoottext{\dag}{From Table 4 and Figure 34 of \citet{bar11}.}
}
\end{table*}

\subsection{Physical properties of the clumps}

\subsubsection{Temperature}
 
In many works, it is assumed that (sub)millimeter clumps/clouds  would have a common temperature (e.g. \citealt{moo04,kirk06,lop11}). This is  a logical assumption  in the case of isolated clouds, athough it is not valid for regions like Gum 31, where different parts of the nebula are affected by different levels of radiation. Then,  we have used the 70\,$\mu$m and 160\,$\mu$m {\em Herschel} images to obtain the dust temperature of the clumps and   of the whole  Gum 31. To construct the flux-ratio map, the 70\,$\mu$m map was smoothed down to the angular resolution of the 160\,$\mu$m map. We divided the intensities of the maps by the recommended color-correction factors described by \citet{ber10}: 1.05 for the 70\,$\mu$m and 1.29 for the 160\,$\mu$m, values derived from a comparison with Planck data. The zero-level of the PACS archive data is unknown, which may cause problems to derive the background emission in each map. Here we adopted the background levels of 0.0023\,Jy/px for 70\,$\mu$m and 0.005\,Jy/px for 160\,$\mu$m found by \citet{preib12}. Then, the color-temperature map  was constructed as the inverse function of the ratio map of Herschel 70 and 160\,$\mu$m color-and-background-corrected maps, i.e., $T_{\rm c}=f^{-1}_{(T)}$ (see details in \citealt{preib12}).

  Assuming a dust emissivity following a power law \hbox{$\kappa_{\nu}$ $\propto$ $\nu^{\beta}$},  being ${\beta}$ the spectral index of the thermal dust emission,   in the optically thin thermal dust emission regime $f_{(T)}$ has the parametric form:
\begin{equation}
\qquad f_{(T)} = \frac{S_{70}}{S_{160}} = \frac{B(70,T)}{B(160,T)} \left( \frac{70}{160} \right) ^{\beta}
\end{equation}
 where $B(70,T)$ and $B(160,T)$   are the Black Body Planck function for a temperature $T$ at  the frequencies 70 $\mu$m and 160 $\mu$m, respectively. The pixel-to-pixel temperature was calculated assuming a typical  value  $\beta$ = 2  (see \citealt{preib12} and references therein) on the whole Gum\,31 region.  The uncertainty in derived dust temperatures using this method was estimated to be about $\sim$ 10  - 15 $\%$ \citep{preib12}.

\begin{figure*}
\centering
\includegraphics[width=480pt]{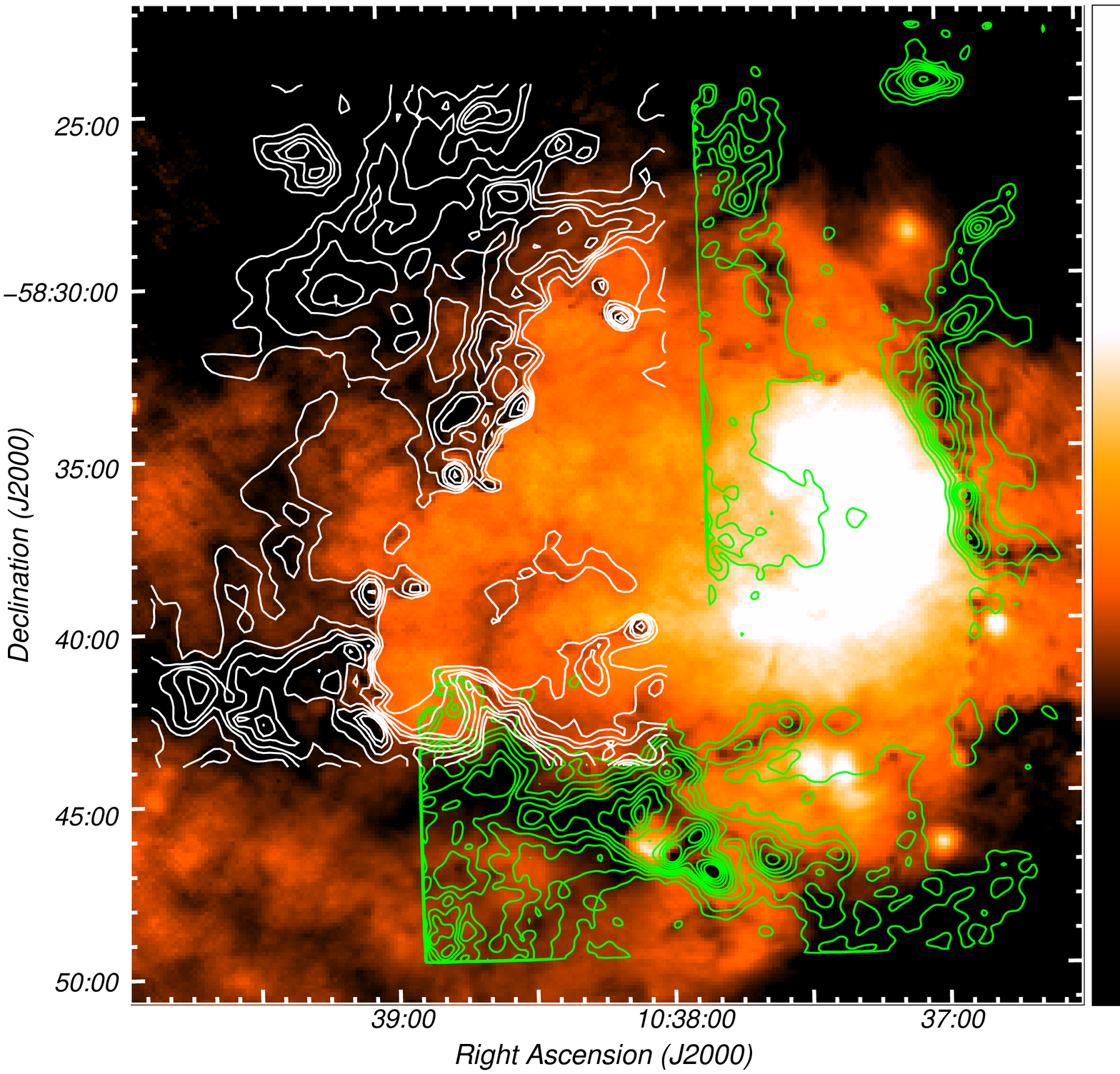}
\caption{Dust  temperature map  (in color scale) derived from {\it Herschel} emission at 70 and 160 $\mu$m. The color-temperature scale is on the right. The 870 $\mu$m emission is shown in green  contours and the 250 $\mu$m emission is shown in white contours. The 870 $\mu$m contours  are the same as in Fig. \ref{870} and the 250 $\mu$m contours  go from  410 \mjyb\ ($\sim$ 4$\sigma$) to 590 \mjyb\ in steps of 50 \mjyb, and from 700 \mjyb\ in steps of 100 \mjyb.   }
\label{fig:tdust}
\end{figure*}

   In Fig.~\ref{fig:tdust} we show the color-temperature map obtained using the method explained above.   The map shows a good correspondence with that obtained in \cite{preib12}  and \citet{ohl13},  indicating temperatures  $\lesssim$ 28 K for  clumps lying at the outer border of the bubble, while clumps projected over the inner region of the bubble (clumps  23, 28, and 60)  show temperatures $\gtrsim$ 28  K, with the exception of clump 49. Inside the bubble, close to the position of the stars of NGC 3324  (where no 870 $\mu$m emission is detected), temperatures are above 35 K. The lowest derived dust  temperature  is 21 K, corresponding to clump 8.  The most remarkable feature  in Fig. \ref{fig:tdust} is the excellent morphological correspondence of the 870 $\mu$m emission with darker regions in the temperature map, which confirms that we are dealing with the  coldest and the  densest molecular gas and dust that were collected behind the ionization front of the \hii\ region bubble.

 Temperatures of the clumps ($T_{\rm dust}$, Col. 10 in Table \ref{tabla-prop}) were simply estimated positioning the center of the fitted  Gaussian on the color temperature map.

\subsubsection{Mass}

 The total (H$_2$ + dust)   mass   of the clumps ($M_{\rm tot}$, Col. 7 in Table \ref{tabla-prop}) was calculated from their integrated 870 $\mu$m emission ($S_{870}$, Col. 4 in Table \ref{tabla-prop}), assuming that the emission is optically thin, using the equation of \citet{hil83}:
\begin{equation}
\qquad M_{(\rm tot)} = R \ \frac{S_{870}\ d^2}{\kappa_{870}\ B_{870}(T_{\rm dust})}
\label{masa}
\end{equation}
where $R$ is the gas-to-dust ratio, $d$ is the distance (adopted as 2.5 kpc), $\kappa_{870}$ is the dust opacity per unit mass at 870 $\mu$m, assumed to be 1.0 cm$^2$ g$^{-1}$  (estimated for dust grains with thin ice mantles in cold clumps; \citealt{osse94}), and $B_{870}(T_{\rm dust})$ is the Planck function for a temperature $T_{\rm dust}$.    For the case of the clumps detected in the {\it Herschel} 250 $\mu$m emission, we used a dust opacity $\kappa_{250}$ = 12.1 cm$^2$ g$^{-1}$, which is consistent with $\kappa_{870}$ = 1.0 cm$^2$ g$^{-1}$ (assuming $\kappa_{\nu}$ $\propto$ $\nu^{2}$).   Adopting $R$ = 186 \citep{jen04,dra07,beu11},  derived masses are in the range from 70 \msun\ to 2800 \msun\ and the total mass is $\sim$ 37600 \msun. This is a lower limit for the total mass in the nebula, since  substantial 870 $\mu$m emission lying outside the  clumps's boundaries was not taken into account (see Fig. \ref{gaussclumps}).  In addition, some of the 870 $\mu$m emission towards the  centre of the nebula (see Sect. 3.1) was probably not detected.  We estimate a total mass uncertainty of about 50 -60 $\%$ arising mostly from a conservative distance  uncertainty of $\sim$ 20 $\%$  and a  dust temperature uncertainty of 15 $\%$.

\subsubsection{Size}

  The effective deconvolved radius of the clumps ($R_{\rm D}$, Col. 7 in Table  \ref{tabla-prop}) was calculated as
\begin{equation}
\qquad  R_{\rm D}\ =\   \sqrt{\frac{\theta_{\rm maj}}{2} \ \times\ \frac{\theta_{\rm min}}{2} }\ ,
\end{equation} 
where  $\theta_{\rm maj}$ and $\theta_{\rm min}$ are the major and minor deconvolved FWHM of the clump (Col. 5  in Table \ref{tabla-prop}). These values  were obtained as:
\begin{equation}
\qquad   \theta_{\rm maj}   =  \sqrt{\ {\rm dx}^2 - \theta_{\rm HPBW}^2}\ \ \textrm{and}
\ \   \theta_{\rm min}   =  \sqrt{\ {\rm dy}^2 - \theta_{\rm HPBW}^2}\ ,
\end{equation}
being dx and dy the major and minor FWHM sizes derived by the  \texttt{Gaussclumps} algorithm, and $\theta_{\rm HPBW}$ the final resolution of the image (20$''$ for 870 $\mu$m and 18$''$ for 250 $\mu$m).  Clump radii are  found to lie between   11\farcs1  and 65\farcs2 (0.16 pc and 0.79 pc, respectively, at a distance of 2.5 kpc) for 870 $\mu$m,  while for 250 $\mu$m clumps radii are between 41\farcs3 to 111\farcs4 (0.5 pc to 1.35 pc at a distance of 2.5 kpc). The difference could reside in that the emission at 250 $\mu$m may be tracing  warmer and less dense  dust than the 870 $\mu$m emission making its emission more extended (see Fig. \ref{glimpse-870}). This could also impact in the determination of the volume densities of the clumps, which are slightly lower than those determined from the 870 $\mu$m emission (see next section).

\subsubsection{Densities}

 The  average volume density of each clump ($n$, Col. 8 in Table \ref{tabla-prop}) was derived, assuming a spherical geometry, as:
\begin{equation}
\qquad  n\ =\  \frac{M_{(\rm tot)}}{4/3\ \pi\   R_{\rm D}^3\ \mu\ m_{\rm H}} 
\end{equation}
where $\mu$ is the mean molecular weight (assumed to be equal to 2.8 after allowance of a relative helium abundance of 25\% by mass), and  $m_{\rm H}$ is the mass of the hydrogen atom. Volume densities of the clumps are between 1.1 $\times$ 10$^3$  cm$^{-3}$ and 2.03 $\times$ 10$^5$  cm$^{-3}$. The  average density  in the sample is   2.9 $\times$ 10$^4$ cm$^{-3}$.

The beam-averaged column density of the clumps ($N_{\rm H_2}$, Col. 9 in Table  \ref{tabla-prop}) was calculated using
\begin{equation}
\qquad  N_{\rm H_2}\   =\ R\ \ \frac{I_{\rm peak}}{\Omega_{\rm beam}\  \kappa_{870}\ \mu\ m_{\rm H}\  B_{870}(T_{\rm dust})}
\label{cmd}
\end{equation} 
\citep{hil83}, where $I_{\rm peak}$ is the 870 $\mu$m emission peak intensity (Col. 3 in Table \ref{tabla-prop})    and $\Omega_{\rm beam}$ is the beam solid angle \hbox{($\pi$ $\theta_{\rm HPBW}^2$ / 4\ ln(2))}. As  we did when determining the mass, we assumed $R$ = 186. Considering a lower limit for the solid angle corresponding to the HPBW size, an intensity peak limit of 3$\sigma$ (150 mJy beam$^{-1}$ for 870 $\mu$m, and 400 mJy beam$^{-1}$ for 250 $\mu$m), and a maximum dust temperature for the clumps of  32 K (see Table \ref{tabla-prop}), the minimum  beam-averaged column densities detectable at 870 $\mu$m and 250 $\mu$m  are 0.63 $\times$ 10$^{22}$ cm$^{-2}$ and  0.32 $\times$ 10$^{22}$ cm$^{-2}$, respectively.

Column densities derived for all the identified clumps are between  0.38 $\times$ 10$^{22}$ cm$^{-2}$  and 10.85  $\times$ 10$^{22}$ cm$^{-2}$. The  average column density of the sample is $\sim$ 2.4 $\times$ 10$^{22}$ cm$^{-2}$, almost twice the value obtained by CNAV08 for the molecular shell associated with Gum 31. This is not surprising given that we are probing the molecular gas intimately associated with the ionization front of the \hii\ region, which is more affected by the expansion of the nebula.  We  keep in mind that this difference could also be due to  different angular  resolutions  and abundance variations.  We also note that volume densities of the 870 $\mu$m clumps are, in average, larger than those of 250 $\mu$m clumps by a factor of $\sim$ 10. The same occurs with column densities, which are larger by a factor of $\sim$ 5. A possible explanation for this difference is  an evolution of the \hii\ region in an inhomogeneous cloud.  The faint 8 $\mu$m, 250 $\mu$m, and 500 $\mu$m emission towards the eastern and northeastern regions of Gum 31 supports this scenario (see Sect. 3.1). On the other hand, as suggested in Sect. 3.3.3, the 250 $\mu$m emission could be tracing  warmer and less dense  dust than the 870 $\mu$m emission. A combination of both factors is also possible.

The physical properties derived for the dust clumps in Gum 31 (radius, density and mass) are similar to those found in IR continuum clumps detected in other Galactic star-forming regions (e.g. \citealt{moo04,lop11,beu11}).

\subsection{Clump mass distribution}

 Using the mass obtained for each clump,   we derived the  clump mass distribution (CMD) in the Gum 31 region. The CMD is  plotted as $d$N/$d$log($M$/M$_{\odot}$) versus mass. The first term is approximated as the number of clumps in each mass interval (N) and the second is approximated  by the logarithmic mass interval, $\Delta {\rm log}$(M/\msun).  The error was determined considering only the statistical Poisson uncertainty, $\sqrt{\Delta {\rm N}}$.   We also  fitted the histograms  with a power law of the form $d$N/$d$log($M$/M$_{\odot}$) $\propto$ $M^{-\alpha}$, where $\alpha$ is the spectral mass index. A  CMD can also be plotted as $d$N/$d$($M$/M$_{\odot}$), which has a power-law dependence $d$N/$d$($M$/M$_{\odot}$) $\propto$ $M^{-\ \Gamma}$,  where  $\Gamma$ = $\alpha$ + 1.

\begin{figure}[h!]
\centering
\includegraphics[width=260pt]{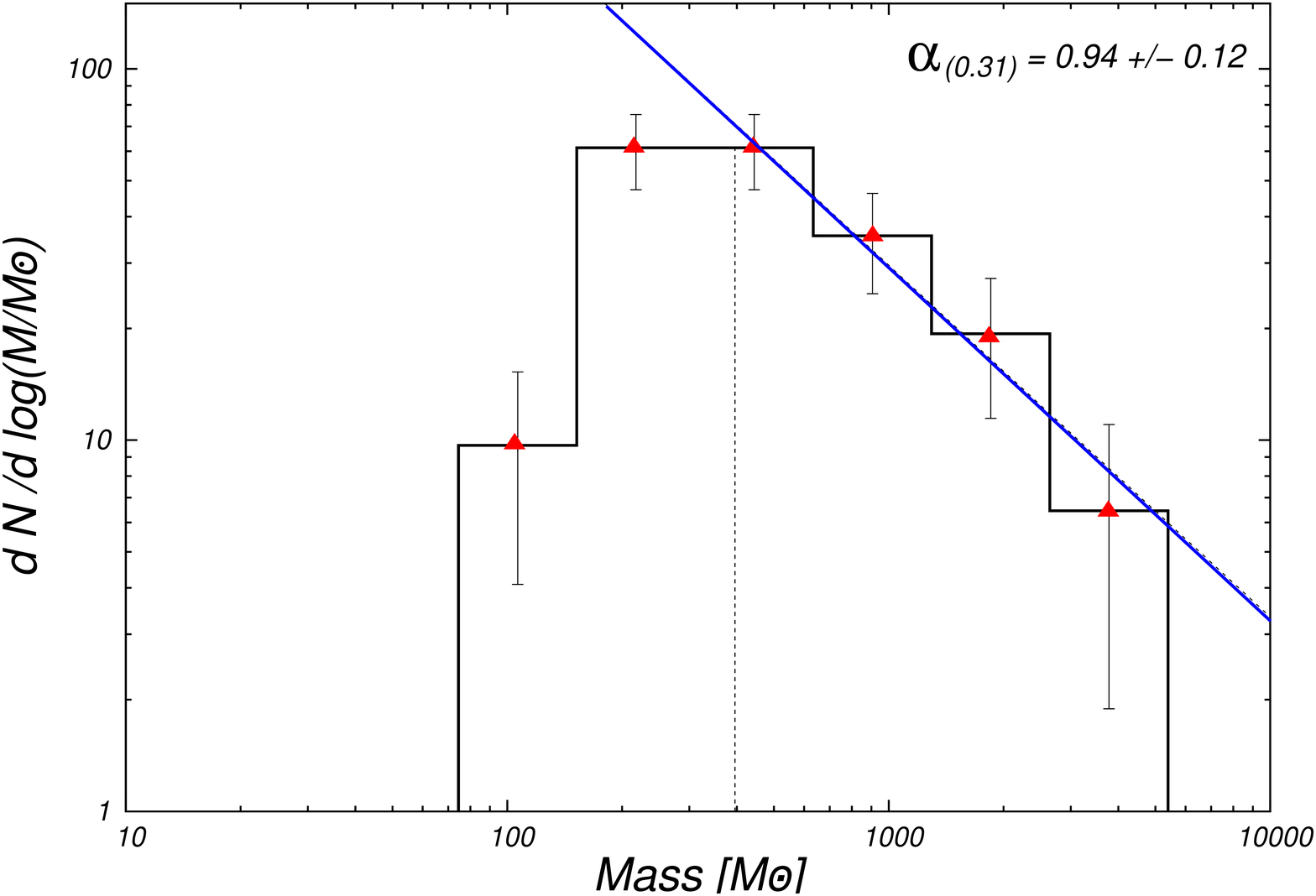}
\caption{ Example of a CMD  for $\Delta {\rm log}$(M/\msun)  = 0.31. Red triangles represent the central mass of each bin and error bars represent the standard deviation of a Poisson distribution ($\sqrt{\Delta {\rm N}}$). Blue lines represent the best fit above the 400 \msun limit (dotted vertical line). }
\label{fig:cmd}
\end{figure}

  To overcome fitting artifacts  due to different bin sizes ($\Delta {\rm log}$(M/\msun)),  we derived  spectral indexes for different bin sizes, ranging from 0.25 to 0.45 in steps of 0.01. For each bin  logarithmic  bin size we fitted the corresponding spectral mass index.   In all the cases, a turnover in the distribution can be seen at bins  between 300 \msun and 500 \msun, which is likely caused by the low-mass incompleteness.        To  make sure that the fit is not affected by incompleteness effects at lower masses, we use a lower cut-off of 400 \msun for the fit.  In Fig. \ref{fig:cmd} we show an example of a  CMD derived for $\Delta {\rm log}$(M/\msun)  = 0.31.  The procedure resulted in 21 fits, obtaining spectral indexes  $\alpha$   in the range   0.44 - 1.35, with  a final  weighted mean spectral index \hbox{$\alpha$ = 0.93 $\pm$ 0.28.}

 Although making a comparison of the spectral mass indexes is a complex task due to the many assumptions authors made, especially when extracting clumps and deriving their masses,    the average  spectral mass index   derived for the dust clumps in Gum 31 is  of the same order of  others derived for molecular clouds from dust continuum (e.g. \citealt{moo04,lop11,beu11,mie12}) and from molecular line CO observations (e.g. \citealt{blitz93,kra98,sim01,hey01,sim06,wong08}).  

\begin{figure*}
\centering
\includegraphics[width=330pt]{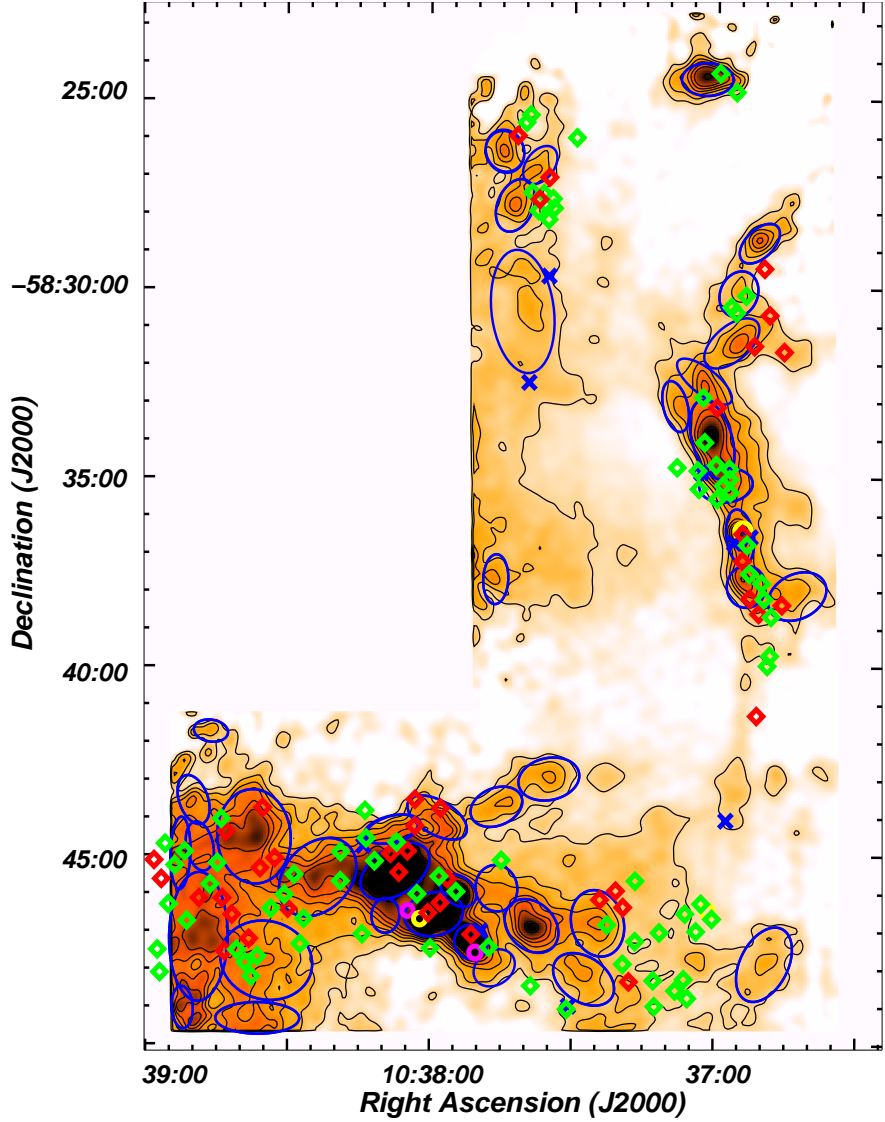}
\caption{ Candidate YSOs projected over the 870 $\mu$m emission. WISE Class I and Class II candidates (OPG13) are marked by red and green diamonds. MSX massive young stellar object (MYSO) and compact \hii\ region candidates (CNAV08)  are marked by magenta and yellow circles, respectively, while 2MASS candidate YSOs (CNAV08) are indicated by blue crosses. Molecular clumps identified in Sect. 3 are indicated  by  blue ellipses. }
\label{fig:ysos}
\end{figure*}

 The slope of a CMD is usually compared to that of the initial stellar mass function (IMF; \citealt{sal55}) as a way to asses a direct relation clump-star \citep{mot98,lada07,lop11}. Equal slopes would indicate that clumps form stars directly. The weighted  average slope derived for Gum 31 is  shallower than  Salpeter's value ($\alpha$ = 1.35 or $\Gamma$ = 2.35) and other indexes between 2.1 and 2.5 in the upper mass regime (above $\sim$ 1 \msun; \citealt{kro01,andre10})    implying that the clumps are not direct progenitors of individual stars. This means that other  processes, such as fragmentation (mainly of the most massive clumps) are necessary to determine the initial stellar masses.  We keep in mind, however, that observations with higher spatial resolution, resolving core structures of $\sim$ 0.1 pc in size, could result in a spectral index of $\sim$2.35 (e. g. \citealt{mot98}).  We also emphasize that the radii of the clumps are between 0.16 pc to as large as 1.35 pc in size (most of them having sizes above 0.4 pc), which are about the size of star-forming regions, rather than individual star-forming cores (typically between 0.01 pc and 0.1 pc in size).  Probably,  smaller clumps (or cores), may not be correctly separated with the resolution of the LABOCA data, although this is just  speculation.  Since these hypothetical smaller cores can not be identified in {\em Herschel} images, higher spatial resolution images (HPBW $<$ 8$''$, equivalent  to  sizes $<$ 0.1 pc at a distance of 2.5 kpc) of the whole nebula would be  useful  to shed some light on this issue.

\begin{figure*}
\centering
\includegraphics[width=480pt]{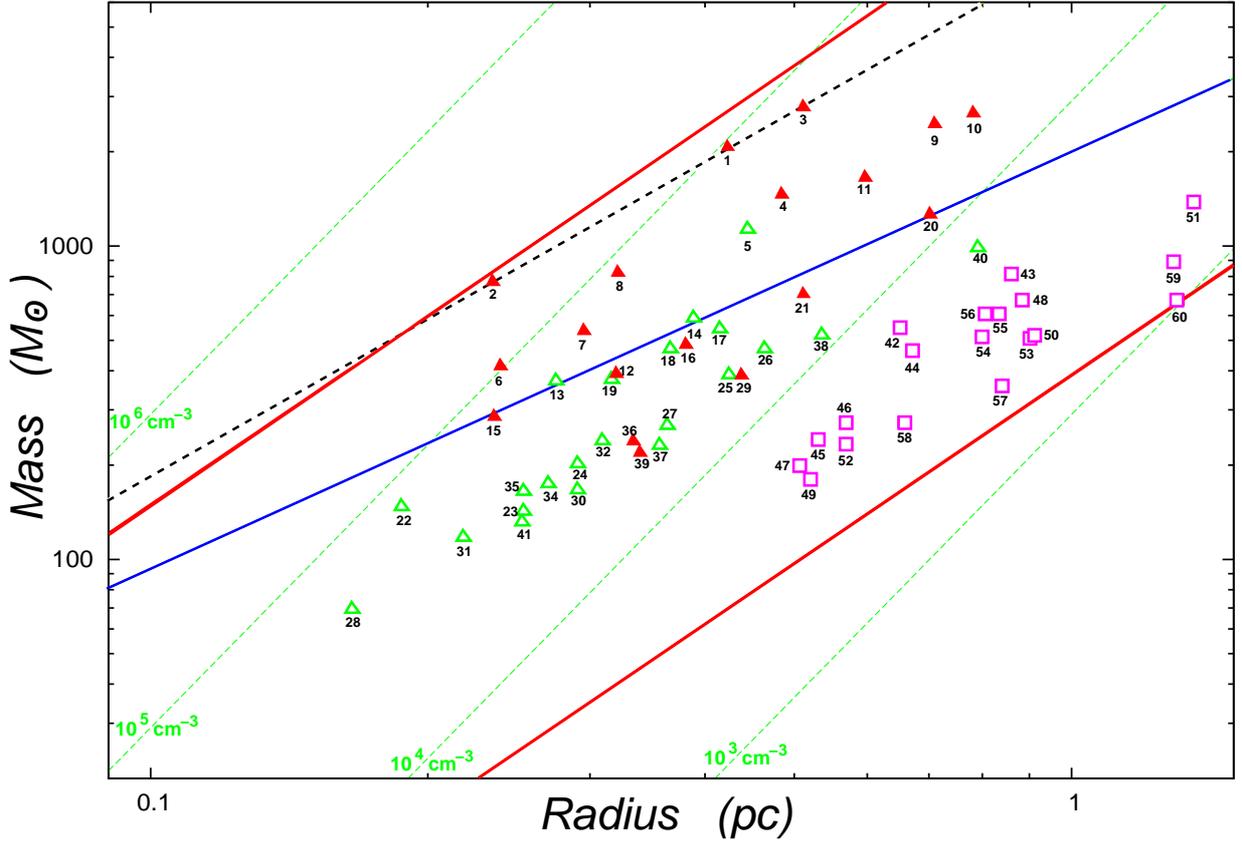}
\caption{  Mass-radius relationship.   Filled red traingles indicate 870 $\mu$m clumps with  candidate YSOs projected inside the clumps's limits determined by \texttt{Gaussclumps}, while green open triangles indicate 870 $\mu$m starless clumps.   Magenta open squares indicate the  250 $\mu$m clumps.  The blue solid line shows  the scaled  mass-radius threshold for high mass star formation  determined by \citet{kau10}.  The rescaled mass-radius relationship derived by \citet{urq13} is shown by the dotted  black line. Upper  and lower solid red lines show surface densities of 1 g cm$^{-2}$ and  0.024  g cm$^{-2}$, respectively.   Dotted green lines indicate volume densities at 10$^3$ cm$^{-3}$,    10$^4$ cm$^{-3}$,  10$^5$ cm$^{-3}$  , and 10$^6$ cm$^{-3}$. }
\label{fig:masaradio}
\end{figure*}

\subsection{Star formation evidence}

 According to \citet{lala03} and \citet{mote03} the radius and mass  required to form stellar clusters are about 0.5 pc to 1 pc and 100 \msun to 1000 \msun, respectively, which are similar to those obtained with  \texttt{Gaussclumps} in Gum 31.  The clumps identified with \texttt{Gaussclumps} in the 870 $\mu$m emission are then  good candidates to form clusters of stars, rather than individual stars.  As pointed out in Sect.1, an exhaustive search of YSO candidates was  carried out by OPG13 using high sensitivity and  high spatial resolution IR data. Although some contamination from the {\bf fore-} and background sources is present, OPG13 considered the WISE-selected sample of YSO candidates to be the more reliable,  finding 661 candidates (207 Class I and 454 Class II sources) of solar  to intermediate mass. Since most of the candidate YSOs detected by  OPG13 and CNAV08 appear projected onto the area mapped with LABOCA,  and the gas and dust in the eastern and northeastern regions of Gum 31 is much less dense than the  southern and western regions,  we will concentrate our analysis only on the 870 $\mu$m emission.

   In Fig. \ref{fig:ysos} we show the location of the WISE YSO candidates reported by OPG13 that  appear projected over the 870 $\mu$m emission. The location of the YSOs were cautiously  determined  using a  composite image of 8.0 $\mu$m, 4.5 $\mu$m, and 3.6 $\mu$m emissions as a reference. We have also included the MSX and 2MASS candidate YSOs identified by CNAV08. From a visual inspection of Fig. \ref{fig:ysos}, we can roughly distinguish five  clusters of YSO candidates  projected onto the submillimeter clumps. Two small compact clusters  are approximately centered  at   \radec\ $\approx$ 10$^h$37$^m$00$^s$, $-$58\gra34\arcmin47\arcsec\  and \radec\ $\approx$ 10$^h$36$^m$54$^s$, $-$58\gra37\arcmin26\arcsec. The first cluster is composed by 10  protostellar members and is projected onto clumps 4, 32, and 39. The second cluster is composed by  13  members and is projected over clumps  7, 12, and 29. In  the southern region of the bubble, three clusters can be discerned. The first one, approximately centered  at  \radec\ $\approx$ 10$^h$38$^m$42$^s$, $-$58\gra45\arcmin45\arcsec, contains approximately 31 members projected over clumps 9, 10, 11, 15, 20, and 34.  The  second one, located at \radec\ $\approx$ 10$^h$38$^m$02$^s$, $-$58\gra45\arcmin42\arcsec, consists of  $\sim$ 29 members projected over clumps 1, 2, 3, 6, 16, 30, and 33,  while the third one is approximately centered  at \radec\ $\approx$ 10$^h$37$^m$13$^s$, $-$58\gra47\arcmin10\arcsec, and is  composed approximately by 18 members projected only over clump 21.

 In Fig. \ref{fig:masaradio} we show the mass-radius relationship for the 41 clumps identified in the 870 $\mu$m emission,     discriminating between clumps that have  candidate YSOs projected inside their limits and clumps without signposts of star formation.  For comparison purposes only, we have also included the 19 clumps identified in the 250 $\mu$m emission. 

 \citet{kau10a,kau10b} and \citet{kau10}  investigated the mass radius relationship of nearby molecular clouds and found the empirical relationship $M\ [{\rm M}_{\odot}] \le 1972 \times\ (R_{\rm eff}\ [{\rm pc}])^{1.33}$\ as the threshold to avoid high mass (M $\sim$ 10 \msun) star formation (hereafter the K$\&$P threshold). Note that when deriving their relationship,  Kauffmann et al. reduced the dust opacities of  \citet{osse94} by a factor of 1.5. Since this correction has not been applied here, we have rescaled the relationship given in the original work (cf. \citealt{urq13}). The rescaling was also calculated considering the opacities and gas-to-dust ratio used in this work ($\kappa_{870}$ = 1.0 cm$^2$ g$^{-1}$ and R = 186;  see Sect. 3.3). The  K$\&$P threshold  is shown in   Fig.~\ref{fig:masaradio}. For the sake of comparison, we also show the empirical mass-size  relationship obtained by \citet{urq13} for intermediate- and high-mass star formation, also rescaled to $\kappa_{870}$ = 1.0 cm$^2$ g$^{-1}$ and R = 186.  The upper and lower red solid lines  in  Fig. \ref{fig:masaradio}   depict the constant surface densities of 1 g cm$^{-2}$   and 0.024 g cm$^{-2}$, respectively.   The surface density of  1 g cm$^{-2}$  (4787 \msun/pc$^2$) was  advocated by \citet{kru08} and \citet{mt03} as criterion to avoid  excessive fragmentation and to allow massive star formation.  The surface density of 0.024 g cm$^{-2}$ represents the average surface density threshold derived by \citet{lada10} (116 \msun pc$^{-2}$) and \citet{hei10} (129 \msun pc$^{-2}$), for efficient star formation.  

 As can be seen from Fig. \ref{fig:masaradio}, all clumps lie above the lower surface density limit of 0.024 g cm$^{-2}$.  Of the 41 clumps identified in the 870 $\mu$m emission, only 15 (37 $\%$) lie above  the K$\&$P limit, with 12 clumps (clumps 1, 2, 3, 4, 6, 7, 8, 9, 10, 15, 11, and 20)  having  candidate YSOs projected inside their limits. Only clumps 5, 13, and 14 are  starless clumps.
 As mentioned in Sect. 3.2, clumps 1 and  3  were also found by \citet{vaz14} (labeled in that work as D1 and D3, respectively) and  were associated by the authors to a dense molecular shell linked to several infrared sources.  Clump 1 is coincident with the MSX sources G286.3747-00.2630 and G286.3773-00.2563, which were classified  as a compact \hii\ region (C\hii) and massive young stellar object (MYSO) candidates, respectively   (sources 20 and 21 in CNAV08). Two WISE Class I and Class II candidates identified by OPG13 are also projected inside the boundaries of this clump (see Fig. \ref{fig:ysos}).     This clump  have an HCO$^{+}$ counterpart (BYF 77b; \citealt{bar11}) which confirms the existence of high density molecular gas. Regarding clump 3, five WISE candidate YSOs identified by OPG13 are projected onto it. This clump also have an HCO$^{+}$ counterpart (BYF 77c).    Clump 2  also have an HCO$^{+}$ counterpart (BYF 77a) and is coincident with the MSX source G286.3579-00.2933 (identified by CNAV08  as  a MYSO candidate), the 2MASS candidate YSO source 10375219-5847133, and two WISE candidate YSOs. For the case of clump 7, a very  crowded spot of candidate YSOs  appear projected inside its limits. The peak intensity of this clump is almost coincident with the position of the MSX C\hii\ candidate G286.1626-00.1877 and the 2MASS sources 10365396-5836293 and 10365749-5836366 (sources 27 and 29 in CNAV08). Three WISE candidate YSOs (OPG13) also appear inside the clump. This clump has an HCO$^+$ counterpart (BYF 70b; \citealt{bar11}). Clumps 6 and 8 are very dense and also have   HCO$^+$ counterparts (BYF 77b and BYF 67, respectively), although only have two and one WISE candidate YSOs projected inside their limits.  Regarding clump 4, it has an  HCO$^{+}$ counterpart (BYF 70a) and  appears projected onto a number of WISE sources although many of them seems to be located at the inner region of the bubble, outside the molecular gas (see Fig. 8 of OPG13). Clumps 9, 10, and 11 are less dense (they do not have  HCO$^{+}$ counterparts), but a considerably number of WISE candidate YSOs are seen projected inside their limits. 

 Several clumps with candidate YSOs projected on them lie below  the K$\&$P mass-radius relation limit, namely: clumps 12, 16, 20, 21, 29, 36, and 39. Probably, the most interesting clumps of this sample are  12 and 39, since they are located  in a region where the star formation activity seems to be very intense (see previous paragraph). Clump 12 has three WISE candidate YSOs projected inside, while clump 39 has six. Both, clumps 12 and 39 have HCO$^{+}$ counterparts (BYF 70b and BYF 70a, respectively).

\subsection{Triggered star formation scenario }

\begin{figure}
\centering
\includegraphics[width=270pt,angle=0]{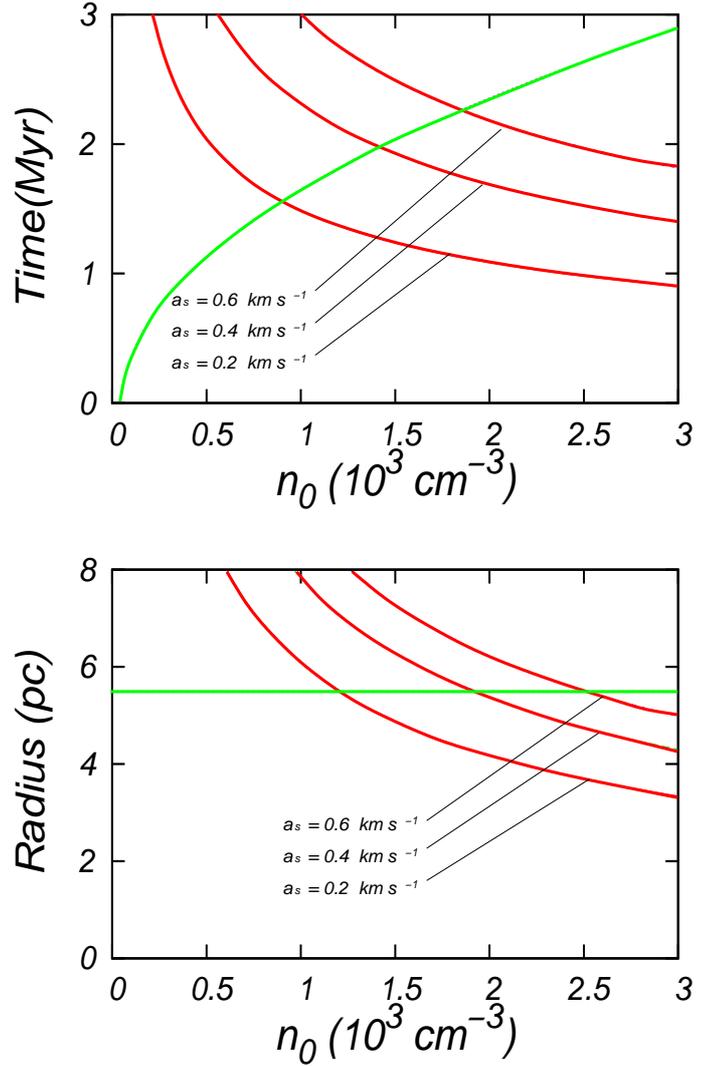}
\caption{ {\it Upper panel:} Fragmentation  time (red curves)  and dynamical time (green curve), for $R$ = 5.5 pc  as  a function of $n_0$ = 0.2, 0.4, and 0.6 \kms. {\it Lower panel:} radius of fragmentation as  a function of $n_0$ for the same values than in the upper panel (red curves). The green curve  represents the radius of the \hii\ region (5.5 pc).  }
\label{fig:fragm}
\end{figure}

As mentioned before, the morphological characteristics of  Gum 31  make this nebula a perfect object to investigate possible scenarios of star formation. To confirm whether the fragmentation of the collected layer of molecular gas and dust have occurred (as suggested  by OPG13 and  in  Sect. 3.4), we use the analytical model of \citet{wi94} for the case of an expanding \hii\ region     which predicts the time at which the fragmentation occurs. We have   followed  the analysis of \citet{zav06}, deriving the time when the  fragmentation may have occurred ($t_{\rm frag}$), and  the size of the \hii region at $t_{\rm frag}$ ($R_{\rm frag}$), given by
\begin{equation}  
\quad t_{\rm frag}\ \  =\ 1.56\  {a_{0.2}}^{7/11}\   n_3^{-5/11}\  {Q({\rm H^o})_{49}^*}^{-1/11} \ \ 10^6 \ \ {\rm yr}
\label{tfrag}                       
\end{equation} 
\begin{equation}  
\quad r_{\rm frag}\ \ =\ 5.8\ {a_{0.2}}^{4/11}\ n_3^{-6/11}\ {Q({\rm H^o})_{49}^*}^{1/11}  \ \ {\rm pc}
\label{rfrag}                       
\end{equation}
where  $a_{0.2}$ is the isothermal sound speed in the compressed layer, in units of 0.2 \kms\ ($a_s/0.2$ \kms),  $n_3$ is surrounding homogeneous infinite medium into which the \hii\ region expands, in units of 1000 cm$^{-3}$ ($n_0/1000$ cm$^{-3}$), and $Q({\rm H^o})_{49}^*$ is the number of ionizing photons, in units of 10$^{49}$ s$^{-1}$ ($Q({\rm H^o})/10^{49}$ s$^{-1}$). In order to determine  $t_{\rm frag}$, in Fig.~\ref{fig:fragm} (upper panel) we  compare Eq. \ref{tfrag} with the standard evolutionary model for expanding \hii\ regions developed by \citet{dw97}, 
\begin{equation}
\quad R\ =\ R_0\   \left[1\ +\ a_{\rm HII}\ \times\  \frac{t_{\rm dyn}}{R_0} \right]^{4/7}
\label{radius}                       
\end{equation}
where $R$ is the observed  radius of the nebula, $a_{\rm HII}$ is the speed of sound in the ionized medium, and     $R_0$ is the radius of the initial  Str{\"o}mgren sphere \citep{s39}  given by 
\begin{equation}
\quad R_0 [n_{\rm 0}, Q({\rm H^o})] = 3.15\ \times\ 10^{-17}\ n_{\rm 0}^{-2/3}\  Q({\rm H^o})^{1/3} \ \ {\rm pc}
\end{equation}
 For the case of NGC\,3324, two O\,6.5V and one O\,8.5 are the main  sources of UV photons (see Sect.1), which provide a total  ionizing flux of  $\sim$ 1.7 $\times$ 10$^{49}$ s$^{-1}$ \citep{sm07}.  Since about 50 $\%$   of the UV photons could be  absorbed by interstellar dust in an \hii region \citep{in01}, we adopt a final value  $Q({\rm H^o})$ $\sim$ 0.85 $\times$ 10$^{49}$ s$^{-1}$.

The upper panel of  Fig. \ref{fig:fragm} shows  $t_{\rm frag}$ and $t_{\rm dyn}$ as function of the initial ambient density $n_0$.  As pointed out in   \citet{wi94},  $a_{s}$ = 0.2  \kms\  is  likely a lower limit for the sound of speed in the collected layer, since both turbulence and extra heating from intense sub-Lyman-continuum photons leaking from the \hii\ region could increase this value. Then, in the following we will adopt a range $a_{s}$ =  0.2 - 0.6 \kms\ for the collected layer. For the ionized medium  we adopt a value  $a_{\rm HII}$ $\simeq $ 10 \kms.  Since $t_{\rm dyn}$ is  required to be higher than $t_{\rm frag}$ for fragmentation,   the initial ambient  density  in the region where Gum 31 was formed must have been in a range $n_0$ $\sim$  950 - 1800  cm$^{-3}$. A slightly larger range ($n_0$ $\sim$ 1100 - 2500 cm$^{-3}$) is derived by comparing $R$ with the radius of fragmentation (Fig.~\ref{fig:fragm}, lower panel).  We have compared these values with the initial ambient density derived using the mass estimates from our  870 $\mu$m data and mass estimates from CNAV08.    If we assume that all the material observed in the nebula next to the ionization front was swept up during its expansion,  an estimate of $n_0$ for Gum 31 can be obtained by assuming that all the gas (either ionized or neutral) was initially located in a sphere of radius  equal to that of the \hii\ region (e.g  \citealt{zav07,ci09,and15}). For the observed radius of the bubble, $R$, we consider the angular radius of $\sim$ 7\farcm 5  derived from the radio continuum emission at 843 MHz (CNAV08) which yields to $R$ = 5.5 pc at a distance of 2.5 kpc. Considering a total molecular mass of 35800 \msun, derived from the 870 $\mu$m  and 250 $\mu$m emission without taking into account the mass of clumps 8, 18, 19, and 35 (see Sect. 3.2)  and an ionized gas mass of 2430 \msun\ (at $d$ = 2.5 kpc) associated to the nebula (3500 \msun at $d$ = 3 kpc; CNAV08), and $R$ =  5.5 pc, we derive an initial ambient density  of  $n_0$ $\sim$ 2200 cm$^{-3}$. This is a lower limit for $n_0$  since a  considerable amount of 870 $\mu$m emission outside the clumps and from the centre of the nebula was not considered (see Sects. 3.2 and  3.2.2). Furthermore, we do not know if the original parental cloud extended to the current radius of the nebula.         The derived  value of $n_0$ is  higher than the density limits required for the fragmentation according to the Whitworth et al.'s scenario, which  indicates that the C$\&$C process might be important  in the collected layers of molecular gas. 

 As pointed out by \citet{zav06}, the estimates presented above are rough, since  the Whitworth et al's model   assumes expansion into an homogeneous medium and many uncertainties remain concerning to the uniformity of the medium where Gum 31 has evolved (see Sect. 3.1). Also, we agree with OPG13 that the presence of cores, bright rims and pillars at the edge of the bubble suggest that other processes, like the interaction of the ionization front with pre-existing molecular condensation (RDI; \citealt{lela94}) or small-scale Jeans gravitational instabilities in the swept-up layers of molecular gas (e.g. \citealt{pom09}) could be also active in some regions.

\section{Summary}

 Using LABOCA 870 $\mu$m continuum emission and IR and optical archival data, we have carried out a study of  the gas and  dust adjacent to the  Gum 31 nebula. The main results can be summarized as follows: 

\begin{enumerate}
 
\item The morphology observed in the 870 $\mu$m emission for Gum 31 is typical  of many Galactic IR  ring nebulae.  The emission is strong towards the border of the \hii\ region and faint towards its center, depicting the typical emission  of a two dimensional projection of {\bf a} spherical bubble (i.e. limb brightening effect). The IR emission distribution coincides with that of the  molecular gas and  suggests that  the cold dust and molecular material have been collected behind the ionization front due to the expansion of the nebula.      The intense optical, 8 $\mu$m, radio continuum, and 870 $\mu$m emissions   at the west  of the powering stars suggest the existence of a PDR  at the interface between the ionized and molecular gas. The high intensity in  radio continuum, optical,  and IR emissions, in the western region of the bubble, as well as the  low intensities at the eastern region suggest that the  \hii\ region is expanding  anisotropically, probably due to the formation of the bubble in an  inhomogeneous parental molecular cloud.     \\  

\item  We have used the \texttt{Gaussclumps} algorithm to identify dust clumps adjacent to the nebula.  After applying some rejection criteria, we were left with 41 clumps in the 870 $\mu$m image. To detect and investigate de dust clumps in the eastern and northestern region of the nebula, which was not covered in the 870 $\mu$m emission data, we made use of an {\em Herschel} image  at 250 $\mu$m and we identified 19 clumps.      \\

\item Adopting a distance of 2.5 kpc, the total (gas + dust) clump masses are in the range from  70 \msun\ to 2800 \msun and the total mass is about 37600 \msun.  The deconvolved linear radius derived for  the clumps are between 0.12 pc and 1.3 pc. Volume and column densities are in the range from  $\sim$ 1.1 $\times$ 10$^3$ cm$^{-3}$  to  $\sim$ 2.04 $\times$ 10$^5$ cm$^{-3}$ and  0.4 $\times$ 10$^{22}$ cm$^{-2}$  to  10.4 $\times$ 10$^{22}$ cm$^{-2}$, respectively.  Dust clump temperatures obtained from high resolution {\em Herschel} data  at 70 $\mu$m and 160 $\mu$m   are between  21 K and  32 K, while dust temperatures at the center of the bubble, close to the ionizing stars of NGC 3324  are as high as  $\sim$ 40 K. \\

\item We have derived the clump mass distributions for Gum 31, approximated as $\Delta$N/$\Delta$log($M$/M$_{\odot}$), using different bin sizes,  and  we fitted a power law  $d$N/$d$log($M$/M$_{\odot}$) $\propto$ $M^{-\alpha}$ for each  bin size. Obtained spectral indexes are in the range  0.44 - 1.35, and the weighted mean  index is $\alpha$ = 0.93 $\pm$ 0.28. The derived spectral index is  of the order of other  indexes derived  in many  works from continuum and molecular line  data for other regions (e.g. \citealt{lop11,wong08}). Although the resolution of LABOCA could not be  sufficient to make a comparison, the difference between the average slope of the CMDs with the slopes of the stellar IMF found in the literature suggests that the clumps are not direct progenitors of single stars. 
\\

\item The mass-radius relationship for the 41  clumps identified in 870 $\mu$m  shows that all the clumps lie above the lower surface density threshold for massive star formation of 0.024 g cm$^{-2}$, while only   15 clumps lie in or above the K$\&$P limit. Of the sample of  clumps lying above the K$\&$P threshold, only three of them do not have candidate YSOs projected inside its their limits. \\

\item   After  analyzing the dynamical age of the \hii\ region and the fragmentation time and radius of the molecular/dust shell surrounding the \hii\ region, we   confirmed that the collect and collapse process  may indeed be important in the collected layers of gas at the edge of the bubble. The presence of YSO candidates detected by OPG13 and CNAV08  projected onto  the 870 $\mu$m emission adjacent to the ionization front is a strong indication in favor of this scenario. We keep in mind, however, that  other process like radiative-driven implosion or small-scale Jeans gravitational instabilities in the swept-up layers of molecular gas,   could also be active in specific regions.  \\

\end{enumerate}

\begin{acknowledgements}

 We very much acknowledge the anonymous referee for her/his helpful comments and suggestions that led to a substantial improvement of this paper.   This project was partially financed by CONICET of Argentina under projects PIP 00356, and  PIP00107,    and  from UNLP, 2012-2014 PPID/G002 and  11/G120. NUD acknowledges C. Kramer for his help running \texttt{Gaussclumps}, and J. Williams for his help running \texttt{Clumpfind}.    VF acknowledges support from ESO-Chile Joint Committee and DIULS. LC acknowledges support from DIULS.   L. G. receives support from the Center of Excellence in Astrophysics and Associated Technologies (PFB-06), CONICYT (Chile) and CSIRO Astronomy and Space Science (Australia).    M.R.  wishes to acknowledge support from CONICYT (CHILE) through FONDECYT grant No1140839.

\end{acknowledgements}

\bibliographystyle{aa}
\bibliography{bibliografia-gum31}

\begin{thebibliography}{93}
\expandafter\ifx\csname natexlab\endcsname\relax\def\natexlab#1{#1}\fi

\bibitem[{{Anderson} {et~al.}(2015){Anderson}, {Deharveng}, {Zavagno},
  {Tremblin}, {Lowe}, {Cunningham}, {Jones}, {Mullins}, \& {Redman}}]{and15}
{Anderson}, L.~D., {Deharveng}, L., {Zavagno}, A., {et~al.} 2015, \apj, 800,
  101

\bibitem[{{Anderson} {et~al.}(2012){Anderson}, {Zavagno}, {Deharveng},
  {Abergel}, {Motte}, {Andr{\'e}}, {Bernard}, {Bontemps}, {Hennemann}, {Hill},
  {Rod{\'o}n}, {Roussel}, \& {Russeil}}]{and12}
{Anderson}, L.~D., {Zavagno}, A., {Deharveng}, L., {et~al.} 2012, \aap, 542,
  A10

\bibitem[{{Andr{\'e}} {et~al.}(2010){Andr{\'e}}, {Men'shchikov}, {Bontemps},
  {K{\"o}nyves}, {Motte}, {Schneider}, {Didelon}, {Minier}, {Saraceno},
  {Ward-Thompson}, {di Francesco}, {White}, {Molinari}, {Testi}, {Abergel},
  {Griffin}, {Henning}, {Royer}, {Mer{\'{\i}}n}, {Vavrek}, {Attard},
  {Arzoumanian}, {Wilson}, {Ade}, {Aussel}, {Baluteau}, {Benedettini},
  {Bernard}, {Blommaert}, {Cambr{\'e}sy}, {Cox}, {di Giorgio}, {Hargrave},
  {Hennemann}, {Huang}, {Kirk}, {Krause}, {Launhardt}, {Leeks}, {Le Pennec},
  {Li}, {Martin}, {Maury}, {Olofsson}, {Omont}, {Peretto}, {Pezzuto}, {Prusti},
  {Roussel}, {Russeil}, {Sauvage}, {Sibthorpe}, {Sicilia-Aguilar}, {Spinoglio},
  {Waelkens}, {Woodcraft}, \& {Zavagno}}]{andre10}
{Andr{\'e}}, P., {Men'shchikov}, A., {Bontemps}, S., {et~al.} 2010, \aap, 518,
  L102

\bibitem[{{Barnes} {et~al.}(2011){Barnes}, {Yonekura}, {Fukui}, {Miller},
  {M{\"u}hlegger}, {Agars}, {Miyamoto}, {Furukawa}, {Papadopoulos}, {Jones},
  {Hernandez}, {O'Dougherty}, \& {Tan}}]{bar11}
{Barnes}, P.~J., {Yonekura}, Y., {Fukui}, Y., {et~al.} 2011, \apjs, 196, 12

\bibitem[{{Baumgardt} {et~al.}(2000){Baumgardt}, {Dettbarn}, \&
  {Wielen}}]{baum00}
{Baumgardt}, H., {Dettbarn}, C., \& {Wielen}, R. 2000, \aaps, 146, 251

\bibitem[{{Beaumont} \& {Williams}(2010)}]{bea10}
{Beaumont}, C.~N. \& {Williams}, J.~P. 2010, \apj, 709, 791

\bibitem[{{Benjamin} {et~al.}(2003){Benjamin}, {Churchwell}, {Babler}, {Bania},
  {Clemens}, {Cohen}, {Dickey}, {Indebetouw}, {Jackson}, {Kobulnicky},
  {Lazarian}, {Marston}, {Mathis}, {Meade}, {Seager}, {Stolovy}, {Watson},
  {Whitney}, {Wolff}, \& {Wolfire}}]{b03}
{Benjamin}, R.~A., {Churchwell}, E., {Babler}, B.~L., {et~al.} 2003, \pasp,
  115, 953

\bibitem[{{Bergin} \& {Tafalla}(2007)}]{ber07}
{Bergin}, E.~A. \& {Tafalla}, M. 2007, \araa, 45, 339

\bibitem[{{Bernard} {et~al.}(2010){Bernard}, {Paradis}, {Marshall}, {Montier},
  {Lagache}, {Paladini}, {Veneziani}, {Brunt}, {Mottram}, {Martin},
  {Ristorcelli}, {Noriega-Crespo}, {Compi{\`e}gne}, {Flagey}, {Anderson},
  {Popescu}, {Tuffs}, {Reach}, {White}, {Benedettini}, {Calzoletti},
  {Digiorgio}, {Faustini}, {Juvela}, {Joblin}, {Joncas}, {Mivilles-Deschenes},
  {Olmi}, {Traficante}, {Piacentini}, {Zavagno}, \& {Molinari}}]{ber10}
{Bernard}, J.-P., {Paradis}, D., {Marshall}, D.~J., {et~al.} 2010, \aap, 518,
  L88

\bibitem[{{Beuther} {et~al.}(2011){Beuther}, {Linz}, {Henning}, {Bik},
  {Wyrowski}, {Schuller}, {Schilke}, {Thorwirth}, \& {Kim}}]{beu11}
{Beuther}, H., {Linz}, H., {Henning}, T., {et~al.} 2011, \aap, 531, A26

\bibitem[{{Blitz}(1993)}]{blitz93}
{Blitz}, L. 1993, in Protostars and Planets III, ed. E.~H. {Levy} \& J.~I.
  {Lunine}, 125--161

\bibitem[{{Bock} {et~al.}(1999){Bock}, {Large}, \& {Sadler}}]{b99}
{Bock}, D.~C.-J., {Large}, M.~I., \& {Sadler}, E.~M. 1999, \aj, 117, 1578

\bibitem[{{Brand} {et~al.}(2011){Brand}, {Massi}, {Zavagno}, {Deharveng}, \&
  {Lefloch}}]{bra11}
{Brand}, J., {Massi}, F., {Zavagno}, A., {Deharveng}, L., \& {Lefloch}, B.
  2011, \aap, 527, A62

\bibitem[{{Cantalupo} {et~al.}(2010){Cantalupo}, {Borrill}, {Jaffe}, {Kisner},
  \& {Stompor}}]{cant10}
{Cantalupo}, C.~M., {Borrill}, J.~D., {Jaffe}, A.~H., {Kisner}, T.~S., \&
  {Stompor}, R. 2010, \apjs, 187, 212

\bibitem[{{Cappa} {et~al.}(2009){Cappa}, {Rubio}, {Mart{\'{\i}}n}, \&
  {Romero}}]{ca09}
{Cappa}, C.~E., {Rubio}, M., {Mart{\'{\i}}n}, M.~C., \& {Romero}, G.~A. 2009,
  \aap, 508, 759

\bibitem[{{Cesarsky} {et~al.}(1996){Cesarsky}, {Lequeux}, {Abergel}, {Perault},
  {Palazzi}, {Madden}, \& {Tran}}]{ces96}
{Cesarsky}, D., {Lequeux}, J., {Abergel}, A., {et~al.} 1996, \aap, 315, L309

\bibitem[{{Churchwell} {et~al.}(2006){Churchwell}, {Povich}, {Allen}, {Taylor},
  {Meade}, {Babler}, {Indebetouw}, {Watson}, {Whitney}, {Wolfire}, {Bania},
  {Benjamin}, {Clemens}, {Cohen}, {Cyganowski}, {Jackson}, {Kobulnicky},
  {Mathis}, {Mercer}, {Stolovy}, {Uzpen}, {Watson}, \& {Wolff}}]{chu06}
{Churchwell}, E., {Povich}, M.~S., {Allen}, D., {et~al.} 2006, \apj, 649, 759

\bibitem[{{Churchwell} {et~al.}(2007){Churchwell}, {Watson}, {Povich},
  {Taylor}, {Babler}, {Meade}, {Benjamin}, {Indebetouw}, \& {Whitney}}]{chu07}
{Churchwell}, E., {Watson}, D.~F., {Povich}, M.~S., {et~al.} 2007, \apj, 670,
  428

\bibitem[{{Cichowolski} {et~al.}(2009){Cichowolski}, {Romero}, {Ortega},
  {Cappa}, \& {Vasquez}}]{ci09}
{Cichowolski}, S., {Romero}, G.~A., {Ortega}, M.~E., {Cappa}, C.~E., \&
  {Vasquez}, J. 2009, \mnras, 394, 900

\bibitem[{{Deharveng} {et~al.}(2008){Deharveng}, {Lefloch}, {Kurtz}, {Nadeau},
  {Pomar{\`e}s}, {Caplan}, \& {Zavagno}}]{deh08}
{Deharveng}, L., {Lefloch}, B., {Kurtz}, S., {et~al.} 2008, \aap, 482, 585

\bibitem[{{Deharveng} {et~al.}(2003){Deharveng}, {Lefloch}, {Zavagno},
  {Caplan}, {Whitworth}, {Nadeau}, \& {Mart{\'{\i}}n}}]{deh03}
{Deharveng}, L., {Lefloch}, B., {Zavagno}, A., {et~al.} 2003, \aap, 408, L25

\bibitem[{{Deharveng} {et~al.}(2010){Deharveng}, {Schuller}, {Anderson},
  {Zavagno}, {Wyrowski}, {Menten}, {Bronfman}, {Testi}, {Walmsley}, \&
  {Wienen}}]{deh10}
{Deharveng}, L., {Schuller}, F., {Anderson}, L.~D., {et~al.} 2010, \aap, 523,
  A6

\bibitem[{{Deharveng} {et~al.}(2012){Deharveng}, {Zavagno}, {Anderson},
  {Motte}, {Abergel}, {Andr{\'e}}, {Bontemps}, {Leleu}, {Roussel}, \&
  {Russeil}}]{deh12}
{Deharveng}, L., {Zavagno}, A., {Anderson}, L.~D., {et~al.} 2012, \aap, 546,
  A74

\bibitem[{{Deharveng} {et~al.}(2009){Deharveng}, {Zavagno}, {Schuller},
  {Caplan}, {Pomar{\`e}s}, \& {De Breuck}}]{deh09}
{Deharveng}, L., {Zavagno}, A., {Schuller}, F., {et~al.} 2009, \aap, 496, 177

\bibitem[{{Draine} {et~al.}(2007){Draine}, {Dale}, {Bendo}, {Gordon}, {Smith},
  {Armus}, {Engelbracht}, {Helou}, {Kennicutt}, {Li}, {Roussel}, {Walter},
  {Calzetti}, {Moustakas}, {Murphy}, {Rieke}, {Bot}, {Hollenbach}, {Sheth}, \&
  {Teplitz}}]{dra07}
{Draine}, B.~T., {Dale}, D.~A., {Bendo}, G., {et~al.} 2007, \apj, 663, 866

\bibitem[{{Duronea} {et~al.}(2014){Duronea}, {Vasquez}, {Romero}, {Cappa},
  {Barb{\'a}}, \& {Bronfman}}]{du14}
{Duronea}, N.~U., {Vasquez}, J., {Romero}, G.~A., {et~al.} 2014, \aap, 565, A30

\bibitem[{{Dyson} \& {Williams}(1997)}]{dw97}
{Dyson}, J.~E. \& {Williams}, D.~A. 1997, {The physics of the interstellar
  medium}, ed. {Dyson, J.~E.~\& Williams, D.~A.}

\bibitem[{{Egan} {et~al.}(1998){Egan}, {Shipman}, {Price}, {Carey}, {Clark}, \&
  {Cohen}}]{egan98}
{Egan}, M.~P., {Shipman}, R.~F., {Price}, S.~D., {et~al.} 1998, \apjl, 494,
  L199

\bibitem[{{Elmegreen} \& {Lada}(1977)}]{elm77}
{Elmegreen}, B.~G. \& {Lada}, C.~J. 1977, \apj, 214, 725

\bibitem[{{Forte}(1976)}]{for76}
{Forte}, J.~C. 1976, \aaps, 25, 271

\bibitem[{{Giannetti} {et~al.}(2014){Giannetti}, {Wyrowski}, {Brand},
  {Csengeri}, {Fontani}, {Walmsley}, {Nguyen Luong}, {Beuther}, {Schuller},
  {G{\"u}sten}, \& {Menten}}]{gia14}
{Giannetti}, A., {Wyrowski}, F., {Brand}, J., {et~al.} 2014, \aap, 570, A65

\bibitem[{{Heiderman} {et~al.}(2010){Heiderman}, {Evans}, {Allen}, {Huard}, \&
  {Heyer}}]{hei10}
{Heiderman}, A., {Evans}, II, N.~J., {Allen}, L.~E., {Huard}, T., \& {Heyer},
  M. 2010, \apj, 723, 1019

\bibitem[{{Hernandez} {et~al.}(2011){Hernandez}, {Tan}, {Caselli}, {Butler},
  {Jim{\'e}nez-Serra}, {Fontani}, \& {Barnes}}]{her11}
{Hernandez}, A.~K., {Tan}, J.~C., {Caselli}, P., {et~al.} 2011, \apj, 738, 11

\bibitem[{{Heyer} {et~al.}(2001){Heyer}, {Carpenter}, \& {Snell}}]{hey01}
{Heyer}, M.~H., {Carpenter}, J.~M., \& {Snell}, R.~L. 2001, \apj, 551, 852

\bibitem[{{Hildebrand}(1983)}]{hil83}
{Hildebrand}, R.~H. 1983, \qjras, 24, 267

\bibitem[{{Hollenbach} \& {Tielens}(1997)}]{ht97}
{Hollenbach}, D.~J. \& {Tielens}, A.~G.~G.~M. 1997, \araa, 35, 179

\bibitem[{{Hosokawa} \& {Inutsuka}(2006)}]{hoso06}
{Hosokawa}, T. \& {Inutsuka}, S.-i. 2006, \apj, 646, 240

\bibitem[{{Inoue}(2001)}]{in01}
{Inoue}, A.~K. 2001, \aj, 122, 1788

\bibitem[{{Jeffers} {et~al.}(1963){Jeffers}, {van den Bos}, \&
  {Greeby}}]{jef63}
{Jeffers}, H.~M., {van den Bos}, W.~H., \& {Greeby}, F.~M. 1963, {Index
  catalogue of visual double stars, 1961.0}

\bibitem[{{Jenkins}(2004)}]{jen04}
{Jenkins}, E.~B. 2004, Origin and Evolution of the Elements, 336

\bibitem[{{Kauffmann} \& {Pillai}(2010)}]{kau10}
{Kauffmann}, J. \& {Pillai}, T. 2010, \apjl, 723, L7

\bibitem[{{Kauffmann} {et~al.}(2010{\natexlab{a}}){Kauffmann}, {Pillai},
  {Shetty}, {Myers}, \& {Goodman}}]{kau10a}
{Kauffmann}, J., {Pillai}, T., {Shetty}, R., {Myers}, P.~C., \& {Goodman},
  A.~A. 2010{\natexlab{a}}, \apj, 712, 1137

\bibitem[{{Kauffmann} {et~al.}(2010{\natexlab{b}}){Kauffmann}, {Pillai},
  {Shetty}, {Myers}, \& {Goodman}}]{kau10b}
{Kauffmann}, J., {Pillai}, T., {Shetty}, R., {Myers}, P.~C., \& {Goodman},
  A.~A. 2010{\natexlab{b}}, \apj, 716, 433

\bibitem[{{Kirk} {et~al.}(2006){Kirk}, {Johnstone}, \& {Di Francesco}}]{kirk06}
{Kirk}, H., {Johnstone}, D., \& {Di Francesco}, J. 2006, \apj, 646, 1009

\bibitem[{{Kov{\'a}cs}(2008)}]{kov08}
{Kov{\'a}cs}, A. 2008, in Society of Photo-Optical Instrumentation Engineers
  (SPIE) Conference Series, Vol. 7020, Society of Photo-Optical Instrumentation
  Engineers (SPIE) Conference Series

\bibitem[{{Kramer} {et~al.}(1998){Kramer}, {Stutzki}, {Rohrig}, \&
  {Corneliussen}}]{kra98}
{Kramer}, C., {Stutzki}, J., {Rohrig}, R., \& {Corneliussen}, U. 1998, \aap,
  329, 249

\bibitem[{{Kroupa}(2001)}]{kro01}
{Kroupa}, P. 2001, \mnras, 322, 231

\bibitem[{{Krumholz} \& {McKee}(2008)}]{kru08}
{Krumholz}, M.~R. \& {McKee}, C.~F. 2008, \nat, 451, 1082

\bibitem[{{Lada} {et~al.}(2007){Lada}, {Alves}, \& {Lombardi}}]{lada07}
{Lada}, C.~J., {Alves}, J.~F., \& {Lombardi}, M. 2007, Protostars and Planets
  V, 3

\bibitem[{{Lada} \& {Lada}(2003)}]{lala03}
{Lada}, C.~J. \& {Lada}, E.~A. 2003, \araa, 41, 57

\bibitem[{{Lada} {et~al.}(2010){Lada}, {Lombardi}, \& {Alves}}]{lada10}
{Lada}, C.~J., {Lombardi}, M., \& {Alves}, J.~F. 2010, \apj, 724, 687

\bibitem[{{Lebouteiller} {et~al.}(2007){Lebouteiller}, {Brandl},
  {Bernard-Salas}, {Devost}, \& {Houck}}]{leb07}
{Lebouteiller}, V., {Brandl}, B., {Bernard-Salas}, J., {Devost}, D., \&
  {Houck}, J.~R. 2007, \apj, 665, 390

\bibitem[{{Lefloch} \& {Lazareff}(1994)}]{lela94}
{Lefloch}, B. \& {Lazareff}, B. 1994, \aap, 289, 559

\bibitem[{{L{\'o}pez} {et~al.}(2011){L{\'o}pez}, {Bronfman}, {Nyman}, {May}, \&
  {Garay}}]{lop11}
{L{\'o}pez}, C., {Bronfman}, L., {Nyman}, L.-{\AA}., {May}, J., \& {Garay}, G.
  2011, \aap, 534, A131

\bibitem[{{Ma{\'{\i}}z-Apell{\'a}niz}
  {et~al.}(2004){Ma{\'{\i}}z-Apell{\'a}niz}, {Walborn}, {Galu{\'e}}, \&
  {Wei}}]{maap04}
{Ma{\'{\i}}z-Apell{\'a}niz}, J., {Walborn}, N.~R., {Galu{\'e}}, H.~{\'A}., \&
  {Wei}, L.~H. 2004, \apjs, 151, 103

\bibitem[{{McKee} \& {Tan}(2003)}]{mt03}
{McKee}, C.~F. \& {Tan}, J.~C. 2003, \apj, 585, 850

\bibitem[{{McLean} {et~al.}(2000){McLean}, {Greene}, {Lattanzi}, \&
  {Pirenne}}]{mcl00}
{McLean}, B.~J., {Greene}, G.~R., {Lattanzi}, M.~G., \& {Pirenne}, B. 2000, in
  Astronomical Society of the Pacific Conference Series, Vol. 216, Astronomical
  Data Analysis Software and Systems IX, ed. {N.~Manset, C.~Veillet, \&
  D.~Crabtree}, 145--+

\bibitem[{{Miettinen}(2012)}]{mie12}
{Miettinen}, O. 2012, \aap, 542, A101

\bibitem[{{Molinari} {et~al.}(2010){Molinari}, {Swinyard}, {Bally}, {Barlow},
  {Bernard}, {Martin}, {Moore}, {Noriega-Crespo}, {Plume}, {Testi}, {Zavagno},
  {Abergel}, {Ali}, {Anderson}, {Andr{\'e}}, {Baluteau}, {Battersby},
  {Beltr{\'a}n}, {Benedettini}, {Billot}, {Blommaert}, {Bontemps}, {Boulanger},
  {Brand}, {Brunt}, {Burton}, {Calzoletti}, {Carey}, {Caselli}, {Cesaroni},
  {Cernicharo}, {Chakrabarti}, {Chrysostomou}, {Cohen}, {Compiegne}, {de
  Bernardis}, {de Gasperis}, {di Giorgio}, {Elia}, {Faustini}, {Flagey},
  {Fukui}, {Fuller}, {Ganga}, {Garcia-Lario}, {Glenn}, {Goldsmith}, {Griffin},
  {Hoare}, {Huang}, {Ikhenaode}, {Joblin}, {Joncas}, {Juvela}, {Kirk},
  {Lagache}, {Li}, {Lim}, {Lord}, {Marengo}, {Marshall}, {Masi}, {Massi},
  {Matsuura}, {Minier}, {Miville-Desch{\^e}nes}, {Montier}, {Morgan}, {Motte},
  {Mottram}, {M{\"u}ller}, {Natoli}, {Neves}, {Olmi}, {Paladini}, {Paradis},
  {Parsons}, {Peretto}, {Pestalozzi}, {Pezzuto}, {Piacentini}, {Piazzo},
  {Polychroni}, {Pomar{\`e}s}, {Popescu}, {Reach}, {Ristorcelli}, {Robitaille},
  {Robitaille}, {Rod{\'o}n}, {Roy}, {Royer}, {Russeil}, {Saraceno}, {Sauvage},
  {Schilke}, {Schisano}, {Schneider}, {Schuller}, {Schulz}, {Sibthorpe},
  {Smith}, {Smith}, {Spinoglio}, {Stamatellos}, {Strafella}, {Stringfellow},
  {Sturm}, {Taylor}, {Thompson}, {Traficante}, {Tuffs}, {Umana}, {Valenziano},
  {Vavrek}, {Veneziani}, {Viti}, {Waelkens}, {Ward-Thompson}, {White},
  {Wilcock}, {Wyrowski}, {Yorke}, \& {Zhang}}]{mol10}
{Molinari}, S., {Swinyard}, B., {Bally}, J., {et~al.} 2010, \aap, 518, L100

\bibitem[{{Mookerjea} {et~al.}(2004){Mookerjea}, {Kramer}, {Nielbock}, \&
  {Nyman}}]{moo04}
{Mookerjea}, B., {Kramer}, C., {Nielbock}, M., \& {Nyman}, L.-{\AA}. 2004,
  \aap, 426, 119

\bibitem[{{Motte} {et~al.}(1998){Motte}, {Andre}, \& {Neri}}]{mot98}
{Motte}, F., {Andre}, P., \& {Neri}, R. 1998, \aap, 336, 150

\bibitem[{{Motte} {et~al.}(2003){Motte}, {Schilke}, \& {Lis}}]{mote03}
{Motte}, F., {Schilke}, P., \& {Lis}, D.~C. 2003, \apj, 582, 277

\bibitem[{{Ohlendorf} {et~al.}(2013){Ohlendorf}, {Preibisch}, {Gaczkowski},
  {Ratzka}, {Ngoumou}, {Roccatagliata}, \& {Grellmann}}]{ohl13}
{Ohlendorf}, H., {Preibisch}, T., {Gaczkowski}, B., {et~al.} 2013, \aap, 552,
  A14 (OPG13)

\bibitem[{{Ossenkopf} \& {Henning}(1994)}]{osse94}
{Ossenkopf}, V. \& {Henning}, T. 1994, \aap, 291, 943

\bibitem[{{Ott} \& {Herschel Science Ground Segment Consortium}(2010)}]{ott10}
{Ott}, S. \& {Herschel Science Ground Segment Consortium}. 2010, in American
  Astronomical Society Meeting Abstracts, Vol. 216, American Astronomical
  Society Meeting Abstracts 216, 413.10

\bibitem[{{Pomar{\`e}s} {et~al.}(2009){Pomar{\`e}s}, {Zavagno}, {Deharveng},
  {Cunningham}, {Jones}, {Kurtz}, {Russeil}, {Caplan}, \&
  {Comer{\'o}n}}]{pom09}
{Pomar{\`e}s}, M., {Zavagno}, A., {Deharveng}, L., {et~al.} 2009, \aap, 494,
  987

\bibitem[{{Povich} {et~al.}(2007){Povich}, {Stone}, {Churchwell}, {Zweibel},
  {Wolfire}, {Babler}, {Indebetouw}, {Meade}, \& {Whitney}}]{pov07}
{Povich}, M.~S., {Stone}, J.~M., {Churchwell}, E., {et~al.} 2007, \apj, 660,
  346

\bibitem[{{Preibisch} {et~al.}(2012){Preibisch}, {Roccatagliata}, {Gaczkowski},
  \& {Ratzka}}]{preib12}
{Preibisch}, T., {Roccatagliata}, V., {Gaczkowski}, B., \& {Ratzka}, T. 2012,
  \aap, 541, A132

\bibitem[{{Rathborne} {et~al.}(2006){Rathborne}, {Jackson}, \&
  {Simon}}]{rath06}
{Rathborne}, J.~M., {Jackson}, J.~M., \& {Simon}, R. 2006, \apj, 641, 389

\bibitem[{{Romero} \& {Cappa}(2009)}]{ro09}
{Romero}, G.~A. \& {Cappa}, C.~E. 2009, \mnras, 395, 2095

\bibitem[{{Salpeter}(1955)}]{sal55}
{Salpeter}, E.~E. 1955, \apj, 121, 161

\bibitem[{{Samal} {et~al.}(2014){Samal}, {Zavagno}, {Deharveng}, {Molinari},
  {Ojha}, {Paradis}, {Tig{\'e}}, {Pandey}, \& {Russeil}}]{sam14}
{Samal}, M.~R., {Zavagno}, A., {Deharveng}, L., {et~al.} 2014, \aap, 566, A122

\bibitem[{{Simon} {et~al.}(2001){Simon}, {Jackson}, {Clemens}, {Bania}, \&
  {Heyer}}]{sim01}
{Simon}, R., {Jackson}, J.~M., {Clemens}, D.~P., {Bania}, T.~M., \& {Heyer},
  M.~H. 2001, \apj, 551, 747

\bibitem[{{Simon} {et~al.}(2006){Simon}, {Rathborne}, {Shah}, {Jackson}, \&
  {Chambers}}]{sim06}
{Simon}, R., {Rathborne}, J.~M., {Shah}, R.~Y., {Jackson}, J.~M., \&
  {Chambers}, E.~T. 2006, \apj, 653, 1325

\bibitem[{{Simpson} {et~al.}(2012){Simpson}, {Povich}, {Kendrew}, {Lintott},
  {Bressert}, {Arvidsson}, {Cyganowski}, {Maddison}, {Schawinski}, {Sherman},
  {Smith}, \& {Wolf-Chase}}]{simp12}
{Simpson}, R.~J., {Povich}, M.~S., {Kendrew}, S., {et~al.} 2012, \mnras, 424,
  2442

\bibitem[{{Siringo} {et~al.}(2007){Siringo}, {Weiss}, {Kreysa}, {Schuller},
  {Kovacs}, {Beelen}, {Esch}, {Gem{\"u}nd}, {Jethava}, {Lundershausen},
  {Menten}, {G{\"u}sten}, {Bertoldi}, {De Breuck}, {Nyman}, {Haller}, \&
  {Beeman}}]{sir07}
{Siringo}, G., {Weiss}, A., {Kreysa}, E., {et~al.} 2007, The Messenger, 129, 2

\bibitem[{{Smith} \& {Brooks}(2007)}]{sm07}
{Smith}, N. \& {Brooks}, K.~J. 2007, \mnras, 379, 1279

\bibitem[{{Str{\"o}mgren}(1939)}]{s39}
{Str{\"o}mgren}, B. 1939, \apj, 89, 526

\bibitem[{{Stutzki} \& {Guesten}(1990)}]{stu90}
{Stutzki}, J. \& {Guesten}, R. 1990, \apj, 356, 513

\bibitem[{{Thompson} {et~al.}(2012){Thompson}, {Urquhart}, {Moore}, \&
  {Morgan}}]{tho12}
{Thompson}, M.~A., {Urquhart}, J.~S., {Moore}, T.~J.~T., \& {Morgan}, L.~K.
  2012, \mnras, 421, 408

\bibitem[{{Traficante} {et~al.}(2011){Traficante}, {Calzoletti}, {Veneziani},
  {Ali}, {de Gasperis}, {di Giorgio}, {Faustini}, {Ikhenaode}, {Molinari},
  {Natoli}, {Pestalozzi}, {Pezzuto}, {Piacentini}, {Piazzo}, {Polenta}, \&
  {Schisano}}]{traf11}
{Traficante}, A., {Calzoletti}, L., {Veneziani}, M., {et~al.} 2011, \mnras,
  416, 2932

\bibitem[{{Urquhart} {et~al.}(2013){Urquhart}, {Moore}, {Schuller}, {Wyrowski},
  {Menten}, {Thompson}, {Csengeri}, {Walmsley}, {Bronfman}, \&
  {K{\"o}nig}}]{urq13}
{Urquhart}, J.~S., {Moore}, T.~J.~T., {Schuller}, F., {et~al.} 2013, \mnras,
  431, 1752

\bibitem[{{Vasquez} {et~al.}(2012){Vasquez}, {Rubio}, {Cappa}, \&
  {Duronea}}]{va12}
{Vasquez}, J., {Rubio}, M., {Cappa}, C.~E., \& {Duronea}, N.~U. 2012, \aap,
  545, A89

\bibitem[{{Vazzano} {et~al.}(2014){Vazzano}, {Cappa}, {Vasquez}, {Rubio}, \&
  {Romero}}]{vaz14}
{Vazzano}, M.~M., {Cappa}, C.~E., {Vasquez}, J., {Rubio}, M., \& {Romero},
  G.~A. 2014, \aap, 570, A109

\bibitem[{{Walborn}(1982)}]{wal82}
{Walborn}, N.~R. 1982, \aj, 87, 1300

\bibitem[{{Watson} {et~al.}(2008){Watson}, {Povich}, {Churchwell}, {Babler},
  {Chunev}, {Hoare}, {Indebetouw}, {Meade}, {Robitaille}, \& {Whitney}}]{wat08}
{Watson}, C., {Povich}, M.~S., {Churchwell}, E.~B., {et~al.} 2008, \apj, 681,
  1341

\bibitem[{{Whitworth} {et~al.}(1994){Whitworth}, {Bhattal}, {Chapman},
  {Disney}, \& {Turner}}]{wi94}
{Whitworth}, A.~P., {Bhattal}, A.~S., {Chapman}, S.~J., {Disney}, M.~J., \&
  {Turner}, J.~A. 1994, \mnras, 268, 291

\bibitem[{{Williams} {et~al.}(2000){Williams}, {Blitz}, \& {McKee}}]{will2000}
{Williams}, J.~P., {Blitz}, L., \& {McKee}, C.~F. 2000, Protostars and Planets
  IV, 97

\bibitem[{{Wong} {et~al.}(2008){Wong}, {Ladd}, {Brisbin}, {Burton}, {Bains},
  {Cunningham}, {Lo}, {Jones}, {Thomas}, {Longmore}, {Vigan}, {Mookerjea},
  {Kramer}, {Fukui}, \& {Kawamura}}]{wong08}
{Wong}, T., {Ladd}, E.~F., {Brisbin}, D., {et~al.} 2008, \mnras, 386, 1069

\bibitem[{{Yonekura} {et~al.}(2005){Yonekura}, {Asayama}, {Kimura}, {Ogawa},
  {Kanai}, {Yamaguchi}, {Barnes}, \& {Fukui}}]{yon05}
{Yonekura}, Y., {Asayama}, S., {Kimura}, K., {et~al.} 2005, \apj, 634, 476

\bibitem[{{Zavagno} {et~al.}(2006){Zavagno}, {Deharveng}, {Comer{\'o}n},
  {Brand}, {Massi}, {Caplan}, \& {Russeil}}]{zav06}
{Zavagno}, A., {Deharveng}, L., {Comer{\'o}n}, F., {et~al.} 2006, \aap, 446,
  171

\bibitem[{{Zavagno} {et~al.}(2007){Zavagno}, {Pomar{\`e}s}, {Deharveng},
  {Hosokawa}, {Russeil}, \& {Caplan}}]{zav07}
{Zavagno}, A., {Pomar{\`e}s}, M., {Deharveng}, L., {et~al.} 2007, \aap, 472,
  835

\bibitem[{{Zavagno} {et~al.}(2010){Zavagno}, {Russeil}, {Motte}, {Anderson},
  {Deharveng}, {Rod{\'o}n}, {Bontemps}, {Abergel}, {Baluteau}, {Sauvage},
  {Andr{\'e}}, {Hill}, \& {White}}]{zav10}
{Zavagno}, A., {Russeil}, D., {Motte}, F., {et~al.} 2010, \aap, 518, L81

\end{thebibliography}
 
\IfFileExists{\jobname.bbl}{}
{\typeout{}
\typeout{****************************************************}
\typeout{****************************************************}
\typeout{** Please run "bibtex \jobname" to optain}
\typeout{** the bibliography and then re-run LaTeX}
\typeout{** twice to fix the references!}
\typeout{****************************************************}
\typeout{****************************************************}
\typeout{}

}

\end{document}